\documentclass[fleqn,usenatbib]{mnras}
\usepackage{newtxtext,newtxmath}
\usepackage[T1]{fontenc}
\usepackage{ae,aecompl}
\usepackage{graphicx}
\usepackage{url}
\newcommand{\jks}{\mbox{$J\!-\!K_{\rm s}$}}
\newcommand{\ks}{\mbox{$K_{\rm s}$}}

\title[Dust production of carbon stars in the MCs]{The mass-loss, expansion velocities and dust production rates of carbon stars in the Magellanic Clouds}
\author[Nanni et al.]{Ambra Nanni$^{1,2}$, Martin A.T. Groenewegen$^3$, Bernhard Aringer$^1$, Stefano Rubele$^{1,4}$, \newauthor Alessandro Bressan$^5$, Jacco Th. van Loon$^6$, Steven R. Goldman$^7$, Martha L. Boyer$^7$
  \\
  $^1$ Dipartimento di Fisica e Astronomia Galileo Galilei,
  Universit\`a di Padova, Vicolo dell'Osservatorio 3, I-35122 Padova, Italy\\
  $^2$ Aix Marseille Univ, CNRS, CNES, LAM, Marseille, France\\
  $^3$ Koninklijke Sterrenwacht van Belgi\"e, Ringlaan 3, B-1180 Brussel, Belgium  \\
  $^4$ Osservatorio Astronomico di Padova, Vicolo dell'Osservatorio 5,
  I-35122 Padova, Italy \\
  $^5$ SISSA, via Bonomea 265, I-34136 Trieste, Italy\\
  $^6$ Lennard-Jones Laboratories, Keele University, ST5 5BG, UK\\
  $^7$ STScI, 3700 San Martin Drive, Baltimore, MD 21218 USA
}
\begin{document}
\label{firstpage}
\date{Accepted .  Received ; in original form }

\pubyear{2019}
\pagerange{\pageref{firstpage}--\pageref{lastpage}}
\maketitle

\begin{abstract}
The properties of carbon stars in the Magellanic Clouds (MCs) and their total dust production rates are predicted by fitting their spectral energy distributions (SED) over pre-computed grids of spectra reprocessed by dust. The grids are calculated as a function of the stellar parameters by consistently following the growth for several dust species in their circumstellar envelopes, coupled with a stationary wind. Dust radiative transfer is computed taking as input the results of the dust growth calculations.
The optical constants for amorphous carbon are selected in order to reproduce different observations in the infrared and optical bands of \textit{Gaia} Data Release 2.
We find a tail of extreme mass-losing carbon stars in the Large Magellanic Cloud (LMC) with low gas-to-dust ratios that is not present in the Small Magellanic Cloud (SMC). 
Typical gas-to-dust ratios are around $700$ for the extreme stars, but they can be down to $\sim160$--$200$ and $\sim100$ for a few sources in the SMC and in the LMC, respectively.
The total dust production rate for the carbon star population is $\sim 1.77\pm 0.45\times10^{-5}$~M$_\odot$~yr$^{-1}$, for the LMC, and $\sim 2.52\pm 0.96 \times 10^{-6}$~M$_\odot$~yr$^{-1}$, for the SMC.
The extreme carbon stars observed with the Atacama Large Millimeter Array  and their wind speed are studied in detail. For the most dust-obscured star in this sample the estimated mass-loss rate is  $\sim 6.3 \times 10^{-5}$~M$_\odot$~yr$^{-1}$.
The grids of spectra are available at: \url{https://ambrananni085.wixsite.com/ambrananni/online-data-1} and included in the SED-fitting python package for fitting evolved stars \url{https://github.com/s-goldman/Dusty-Evolved-Star-Kit}.
\end{abstract}
\section{Introduction}
\label{introduction}
During the thermally pulsing asymptotic giant branch (TP-AGB) phase, low- and intermediate-massive stars lose mass at high rates, between $\sim10^{-7}$ and $\sim10^{-4}$~M$_\odot$~yr$^{-1}$, enriching in metals and dust the insterstellar medium of galaxies. 
The dense environment of the circumstellar envelopes (CSEs) of TP-AGB stars represents the ideal site for dust condensation.
Stellar pulsation triggers shock waves that lift the gas above the stellar surface where the temperature is low enough to allow solid particles to form and to accelerate the outflow if sufficient momentum is transferred \citep{Hoefner18}.
TP-AGB stars have been shown to be important dust producers in local and maybe also in high-redshift galaxies, under some specific assumptions for the star formation history \citep{Gehrz89, Valiante09, Dwek11}.

Furthermore, dust grains deeply affect the spectra and colours of
TP-AGB stars, because of their ability to reprocess the stellar radiation by absorbing and scattering photons from the stellar photosphere. The emerging spectra 
depend on the chemistry, structure and size of dust grains, which determine their optical properties, and on the amount of dust
condensed in the CSEs that produces different degrees of obscuration. 
The dust chemistry is mainly affected by the number of
carbon over the number of oxygen atoms ($C/O$) in the stellar atmosphere while 
the amount of dust condensed depends on the stellar parameters.

For metallicities lower than solar, characteristic of the Magellanic Clouds (MCs), a large fraction
of TP-AGB stars evolves through the carbon phase ($C/O>1$). Therefore, carbon stars are
extremely relevant for the interpretation of Near and Mid infrared (NIR and MIR) colours in these galaxies.
Among the dust species formed around carbon stars, amorphous carbon (amC) is usually the dominant opacity source shaping the spectral energy distribution (SED) of these stars.
However, several optical data sets for amC dust, yielding very different spectra, are available in the literature \citep{Hanner88,Rouleau91,Zubko96,Jaeger98}. 
The optical data set for carbon dust employed in the radiative transfer calculations, together with the grain size, can be constrained by reproducing both the infrared and optical colours \citep{Nanni16, Nanni18, Nanni19}. Besides, the effect of employing different optical data sets for amC dust have been tested by \citet{Andersen99} in the different context of hydrodynamical simulations.

We use radiative transfer calculations to perform the SED fitting of dust-enshrouded TP-AGB stars in order to estimate their
current dust production rates (DPRs), mass-loss rates\footnote{With the expression ``mass-loss rate'' we always refer to the gas mass-loss rate.} and luminosities. 
For TP-AGB stars in the MCs and in other nearby galaxies several authors have adopted this method \citep{vanLoon99,vanLoon05,vanLoon06_2,Groenewegen07, Groenewegen09, Srinivasan11, Gullieuszik12, Boyer12, Riebel12, Matsuura09, Matsuura13, Srinivasan16,Goldman17,Goldman18,Nanni18, Groenewegen18} and different grids of spectra are available in the literature \citep{Groenewegen06,Srinivasan11, Nanni18}.
The main shortcoming of most of the grids of spectra is that the radiative transfer calculations are based on several assumptions concerning the dust chemistry, the dust condensation temperature (which is usually fixed), the radial density profile, the data set of optical constants for the dust and the grain size distribution \citep{Groenewegen09, Srinivasan11}. 
Furthermore, in order to estimate the dust production and mass-loss rates the outflow expansion velocity and gas-to-dust ratio usually need to be assumed. In only few works the measured wind speeds are used to estimate the gas-to-dust ratios \citep{Marshall04, Groenewegen_etal16, Goldman17}.
All these assumptions affect the final estimates of the dust production and mass-loss rates. The estimates from various authors for the MCs can
differ a lot from each other \citep{Boyer12,Matsuura13,Srinivasan11, Srinivasan16,Nanni18}. In addition, it has recently been shown that different choices of the optical data sets for dust produce relevant variations in the estimated mass-loss rates \citep{Srinivasan11,Nanni18, Groenewegen18}.

In \citet{Nanni18} we adopted a new approach for estimating the DPRs in the SMC by computing the grids of spectra reprocessed by dust by computing dust growth for several dust species, coupled with a stationary wind, as a function of the stellar parameters.
This approach allows us to consistently calculate the dust chemistry, the dust condensation temperature, the dust-density profile, the outflow expansion velocity and gas-to-dust ratio for each set of input quantities.
Moreover, we select the combinations of optical constants and grain sizes which best reproduce most of the infrared and optical colours from the \textit{Gaia} data release 2 (DR2) of carbon stars in the MCs \citep{Nanni16, Nanni19}. 
In this work, the aforementioned grids of spectra are employed to provide new estimates of the total DPR, mass-loss rates, luminosities and dust content of carbon stars in the MCs.
\section{Model and grid parameters}\label{model}
The same description of dust growth, wind dynamics and radiative transfer through the CSEs employed for the SED fitting of the carbon stars in the SMC \citep{Nanni18} is adopted here for the calculation of our grids of spectra. 
In this framework, the growth of various dust species is coupled with a stationary wind in spherical symmetry, as discussed in \citet{Nanni13, Nanni14}, which is a revised version of the description by \citet{FG06}.
Various authors have adopted the original scheme by \citet{FG06} for calculating dust condensation along stellar evolutionary tracks \citep{Ventura12,Ventura14, Ventura16,Dellagli15b, Dellagli15a}.
The input quantities of the code are the stellar parameters: (a) luminosity, $L$, (b) effective temperature, $T_{\rm eff}$ and corresponding photospheric spectrum,  (c) current stellar mass, $M$, (d) element abundances in the atmosphere, (e) mass-loss rate, $\dot{M}$. The other input quantities are the seed particle abundance, $\epsilon_{\rm s, C}$, which affects the grain size \citep{Nanni16}, and the set of optical constants for the different dust species. 
The quantity $\epsilon_{\rm s, C}$ is set to be proportional to the carbon excess \citep{Nanni13, Nanni14, Nanni16, Nanni18,Nanni19},
\begin{equation}\label{seed_abundance}
\epsilon_{\rm s, C}\propto\epsilon_{\rm s}\times(C-O),
\end{equation}
where $\epsilon_{\rm s}$ is a model parameter (see also Table~\ref{opt}).
By adopting this relation, we implicitly assume that seed nuclei are composed by carbonaceous material. The dependence of the seed particle abundance with the other stellar parameters, as well as their composition, is unknown. 
It is thus not possible to exclude that metal carbides such as TiC might be the main constituent of the initial seeds, as suggested by \citet{vanLoon08}. In this case, the number of seed nuclei would be proportional to the initial metallicity.
The analysis presented here will not change if the same grain size is obtained in the calculations, independently of the chemical composition of the seed nuclei.

For each dust species $i$ the grain growth is given by the balance between the accretion ($J^{\rm gr}_{i}$) and destruction rates ($J^{\rm dec}_{i}$):
\begin{equation}\label{dadt}
 \frac{da_{i}}{dt}=V_{\rm 0,i} (J^{\rm gr}_{i}-J^{\rm dec}_{i}),
\end{equation}
where $V_{\rm 0,i}$ is the volume of the monomer of dust. Below a certain temperature that depends on the dust species and on the efficiency of different destruction processes, $J^{\rm gr}_i-J^{\rm dec}_i$ becomes $>0$ and the growth term dominates. 
The quantity $J^{\rm gr}_{i}$ is defined as the minimum growth rate of all the molecular species $j$ that are involved in grain growth:
\begin{equation}\label{Jgr}
    J^{\rm gr}_{i}=\alpha_i n_j v_{{\rm th},j},
\end{equation}
where $\alpha_i$ is the sticking coefficient and $n_j$ and $v_{{\rm th},j}$ are the number density of the species $j$ in the gas phase and its thermal velocity, respectively. The sticking coefficient is assumed to be the same for all the gas species, and represents the probability that
a molecule sticks onto the grain surface when a collision occurs.
The dust species included in the calculations are amorphous carbon (amC), silicon carbide (SiC) and metallic iron. 
The value of the sticking coefficient adopted is $\alpha_i=1$ for all the dust species considered. We also assume that amC dust can only grow when the gas temperature is $\leq 1100$ K and that no destruction is occurring \citep{Cherchneff92,FG06}. 

The momentum equation is of the outflow (in spherical symmetry) is:
\begin{equation}\label{velocity}
v \frac{dv}{dt}=-\frac{G M}{r^2}(1-\Gamma),
\end{equation}
where $\Gamma$ represents the ratio between the radiation pressure and the pull of the gravity \citep[See][for all the details]{Nanni13}.
The radiation pressure increases when dust is formed, and if the momentum transferred is large enough the outflow is accelerated.
The drift velocity between the dust and the gas is neglected. 
The gas and dust density profile are computed from equation \ref{velocity}. For the gas we have:
\begin{equation}\label{dens}
\rho=\frac{\dot{M}}{4\upi r^2 v}.
\end{equation}
The density directly affects the grain growth through the term $n_j$ in equations \ref{dadt} and \ref{Jgr}.
The dust density profile is derived by combining equation \ref{dens} with the amount of dust condensed at each time-step (equation \ref{dadt}).

The gas temperature profile is given by:
\begin{equation}\label{gas_temp}
T_{\rm gas}(r)^4=T_{\rm eff}^4\left[W(r)+\frac{3}{4}\tau_{\rm L}\right],
\end{equation}
where $W(r)$ is the dilution term,
$W(r)=\frac{1}{2}\left[1-\sqrt{1-\left(\frac{R_*}{r}\right)^2}\right]$,
and $\tau_{\rm L}$ is obtained by integrating the equation:
\begin{equation}\label{tauL}
\frac{d\tau_{\rm L}}{dr}=-\rho\kappa \left(\frac{R_*}{r}\right)^2, 
\end{equation}
where $\kappa$ represents the opacity of the medium computed as in \citet{Nanni13}, and $R_*$ is the stellar radius.

The system of equations is solved by integrating equations \ref{dadt} for each dust species, \ref{velocity} and \ref{tauL}.
The initial grain size is assumed to be $a_0=10^{-3}$~$\mu$m, while the initial expansion velocity of the outflow is $v_{\rm i}= 4$~km~s$^{-1}$. If the outflow is not accelerated, dust condenses passively in the CSEs, and the value of the expansion velocity is assumed to be constant and equal to $v_{\rm i}$. The value of $v_{\rm i}$ is a model parameter that is consistent with the lower expansion velocity observed for carbon stars in the Galaxy \citep{Schoier13, Ramstedt14, Danilovich15}.

In Table~\ref{opt} the combinations of optical data sets and $\epsilon_{\rm s}$ employed in our calculations are listed.
The range of amC dust grain size obtained are also mentioned.
These combinations simultaneously reproduce the main infrared colour--colour diagrams for carbon stars in the SMC \citep{Nanni16} and their SEDs \citep{Nanni18}. In addition, as presented in \citet{Nanni19}, these optical constants reproduce the trends obtained by combining 2MASS and \textit{Gaia} DR2 photometry for the carbon stars in the LMC, introduced by \citet{Lebzelter18}.
The optical data sets for SiC and metallic iron are taken from \citet{Pegourie88} and \citet{Leksina67}, respectively.
Our dust growth code is coupled with a radiative transfer code \textsc{more of dusty} \citep{Groenewegen12}, based on \textsc{dusty} \citep{Ivezic97}, to compute the spectra (and colours) reprocessed by dust.
The radiative transfer code takes as input $T_{\rm eff}$ and the corresponding photospheric spectrum, and some of the quantities calculated from the dust growth code. 
These quantities are (a) the average scattering and absorption efficiencies ($\bar{Q}_{\rm sca}$, $\bar{Q}_{\rm abs}$), (b) the dust-density profile, $\rho_{\rm d}(r)$, that is computed from equations \ref{dens} and \ref{dadt}, (c) the optical depth at a given wavelength ($\tau_\lambda$) and d) the dust temperature at the inner boundary of the dust condensation zone.
The dust absorption and scattering coefficients for each of the dust species, $i$, are computed for spherical grains by means of the Mie code \textsc{bhmie} by \citet{Bohren83}. The final $\bar{Q}_{\rm sca}$, $\bar{Q}_{\rm abs}$, as well as all the other quantities such as $\tau_\lambda$, are computed for the consistently calculated dust mixture \citep[see also][]{Nanni18, Nanni19}.
All the spectra obtained are normalised to the total luminosity \citep{Ivezic99}.

The optical depth is computed as:
\begin{equation}\label{tau_lambda}
\tau_\lambda=\frac{3\dot{M}}{4}\int_{R_{\rm c}}^{\infty} \sum_i \frac{Q_{\rm ext, i}(\lambda,a_i)}{a_i \rho_{\rm i}}\frac{\delta_i}{r^2 v}dr,
\end{equation}
where $R_{\rm c}$ is the condensation radius of the first dust species condensed (SiC) given in stellar radii R$_*$, and $Q_{\rm ext, i}$, $\delta_i$, $\rho_i$, are the extinction efficiency, the dust-to-gas ratio and the density of the dust species $i$, respectively.

The DPR of the individual stars is derived from equation \ref{tau_lambda} once $\tau_\lambda$ is constrained from the SED fitting:
\begin{equation}\label{dpr_appr}
\dot{M}_{\rm dust}\propto \tau_\lambda v.
\end{equation}
From this equation it is possible to see how the value of the total mass-loss rate is proportional to the expansion velocity and to the dust-to-gas (or gas-to-dust) ratio, for a given $\tau_\lambda$.

The grids of spectra are computed for two metallicity values representative of the carbon stars in the SMC and in the LMC ($Z=0.004, 0.006$). For the SMC the metallicity value is taken from \citet{Rubele18}, while for the LMC carbon stars we select the typical value derived from \textsc{trilegal} simulations \citep{Girardi05} based on the star formation history derived by \citet{Harris09} (Marigo, private communication). The same reference value for the LMC has been assumed in other works \citep{Groenewegen_etal16, Lebzelter18}.
For the $Z=0.004$ the spectra have been calculated by employing a denser sampling of dust density profile employed in the radiative transfer calculations with respect to \citet{Nanni18}. 
With respect to the spectra in \citet{Nanni18} the grids at $Z=0.004$ include higher values of the carbon excess, $C_{\rm ex}=8.7, 9$, where  $C_{\rm ex}=\log{(C-O)}+12$. 
The range of stellar parameters is provided in Table~\ref{Table:grid}. Scaled solar abundances of the elements in the atmosphere (excluding carbon) are adopted. The range of values selected for $T_{\rm eff}$ is between 2500 and 3600 K. Higher effective temperatures that are not typical of carbon stars have been excluded. 
For each of the combinations of stellar parameters the photospheric spectrum is interpolated in the values of $T_{\rm eff}$ and in $C/O$ between the ones available in the \textsc{comarcs} grid \citep{Aringer09,Aringer16}. 
A metallicity of $Z\sim 0.005$ is selected for the photospheric spectra in the \textsc{comarcs} grid, consistent with the value adopted in our calculations.
The spectra are computed for CSEs which provide $10^{-3}\le \tau_1\le 60$. 
The combinations of the input stellar parameters in Table~\ref{Table:grid} yield, for each metallicity value, $\approx 85500$ spectra for the J1000 optical data set and $\approx 91000$ for the H11.
\begin{table*}
\begin{center}
\caption{Combination of optical data sets and seed particle abundances selected on the basis of \citet{Nanni16} and \citet{Nanni19}. The corresponding grain sizes are obtained from the SED fitting procedure presented in this work.}
\label{opt}
\begin{tabular}{c c c c c}
\hline
Optical data set  & $\rho_{\rm d, amC}$ [g cm$^{-3}$]& $\log(\epsilon_{\rm s})$ & amC grain size [$\mu$m] &  Denomination \\
\hline
\citet{Jaeger98} (T=1000 $^\circ$C) & 1.988 &$-12$   & up to $\sim 0.08$ &J1000 \\
\citet{Hanner88} &  1.85 & $-11$ & up to $\sim 0.04$  &H11 \\
\hline
\end{tabular}
\end{center}
\end{table*}

\begin{table}
\begin{center}
\caption{Input stellar parameters and spacing for the grids of models.}
\label{Table:grid}
\begin{tabular}{l l l}
\hline
Parameter  &  Range/values  & spacing \\
\hline
$\log(L/L_\odot)$  & $[3.2,\,4]$  & 0.1   \\
                   & $[4.0,\, 4.4]$  & 0.05  \\
$\log(\dot{M}/M_{\odot} {\rm yr}^{-1})$    & $[-7,\, -5]$  & 0.1  \\
                   &  $[-5.0,\, -4.4]$ &  0.05  \\
$T_{\rm eff}$/K    &  $[2500,\,3600]$ & 100   \\
$M/M_{\odot}$     & 0.8, 1.5, 3  & \\
Z &   0.004, 0.006 &  \\
C$_{\rm ex}$ & 8.0, 8.2, 8.5, 8.7, 9.0  & \\
C/O for Z=0.004 & 1.65, 2, 3, 4.3, 7.5  & \\
C/O for Z=0.006 & 1.4, 1.6, 2.3, 3.1, 5.2  & \\
\hline
\end{tabular}
\end{center}
\end{table}
\section{Available observations of carbon stars}\label{sec:obs}
The carbon stars for the SMC and LMC are selected from the catalogues by \citet{Srinivasan16}, based on the one by \citet{Boyer11}, and \citet{Riebel12}, respectively. 
The catalogue by \citet{Srinivasan16} includes the classification for 81 sources observed by the Spitzer's Infrared Spectrograph (IRS) and classified by \citet{Ruffle15}. 
Recently, \citet{Jones17} and \citet{Groenewegen18} have studied and classified the IRS spectra of a sample of stars.
In the catalogue by \citet{Groenewegen18} carbon stars of both the SMC and LMC are included, while in \citet{Jones17} only LMC sources have been considered.
In addition to that, medium-resolution optical spectra have been obtained and classified by \citet{Boyer15} for 273 sources in the SMC bar and for 3791 stars in the LMC. These spectra have been obtained by means of the AAOmega/2dF multi-object spectrograph for the SMC \citep{Lewis02, Saunders04, Sharp06}, and by the Hydra-CTIO multi-fiber spectrograph for the LMC \citep{Barden98}.

The stars in the catalogues by \citet{Riebel12} and \citet{Srinivasan16} are cross-matched within $1^{\prime\prime}$ of those from \citet{Jones17}, and of the sample studied by \citet{Groenewegen18}. The spectral classification by \citet{Boyer15} has been added as an additional information in the catalogues. We also include in the analysis $11$ carbon stars in the LMC classified by \citet{Boyer15} that were not included in the other catalogues considered.
If the classification is not consistent between the catalogues considered, the designation by \citet{Groenewegen18} is adopted.  
The only exception is represented by HV 942 that is classified as an R Coronae Borealis star by \citet{Jones17} and that is excluded from the sample because of its SED. 
The star J012606.02--720921.0 is excluded from our sample, since this source is characterized by observed photometry that is not compatible with the SED from an AGB star \citep{Srinivasan16, Nanni18}.

In cases where the spectra are not available, carbon stars are selected on the basis of the photometric classification contained in \citet{Riebel12} and \citet{Srinivasan16} which follows the criteria described in \citet{cioni06a} and \citet{blum06}. 
According to this classification, carbon-rich AGB candidates are selected on the basis of their location on the $\ks$ versus $\jks$ colour--magnitude diagram (CMD), while ``extreme'' TP-AGB (X-) stars are selected on the basis of the $J - [3.6]$ colour. In case the $J$-band is not observed, the $[3.6]-[8.0]$ colour is considered. 
If no spectral classification is available, X-stars are assumed to be carbon-rich, even though some OH/IR stars are expected among them \citep{vanLoon97,vanLoon98,Trams99}.
The catalogue by \citet{Srinivasan16} includes two additional class of stars called ``anomalous'' AGB stars \citep[aAGBs;][]{Boyer15} and far-infrared (FIR) sources. The large majority of aAGBs are classified as oxygen-rich according to the photometric classification \citep{Boyer11},  
but about half of them are expected to be carbon rich according to their spectral classification \citep{Boyer15}. The catalogue by \citet{Srinivasan16} includes 17 FIR sources over the 360 included in the catalogue by \citet{Boyer11}, that have been identified as evolved stars. In \citet{Riebel12} stars are instead simply classified on the basis of the photometry selection of \citet{Cioni06} and \citet{blum06}.
In order to have comparable photometric classification in the two catalogues, we include in our analysis the aAGBs and FIR sources classified as carbon stars on the basis of their spectra \citep{Ruffle15, Boyer15} or of their NIR colours \citep{Cioni06}. 
A fraction of carbon rich aAGBs is expected to be excluded by our study according to our selection criteria. However, the amount of dust produced by those stars is expected to be negligible \citep{Srinivasan16, Nanni18}.
All the photometrically selected C- or X-stars that are not classified as carbon on the basis of their IRS or optical spectra are excluded from the catalogues. We instead include those stars photometrically selected as oxygen rich that have been shown to be carbon on the basis of their spectra.

The catalogues by \citet{Riebel12} and \citet{Srinivasan16} do not include 9 possible evolved stars surrounded by cold dust ($<50$ K) \citep{Jones15}. We do not study these objects, since our theoretical approach does not predict a large amount of cold dust able to explain the emission observed in the Herschel bands, preventing a good estimate of their DPRs.

As far as possible contamination in the catalogue are concerned, we expect to have a negligible fraction of young stellar objects in our analysis that can be mistaken for dust-rich evolved stars, as also shown in Fig. 21 of \citet{Srinivasan16}.

The main references for the photometry used in the aforementioned works are summarized in Table~\ref{photometry}.
For the sample selected from \citet{Groenewegen18} the SED fitting is performed by employing the photometry provided in Table~\ref{photometry}.
In case multiple entries are available for the same filter, the average value is considered.
For the remaining sources the photometry from the catalogues by \citet{Srinivasan16} and \citet{Riebel12} is fitted. The errors of the photometric fluxes in these two catalogues take into account the effect of variability from the U to the $\ks$-band, by adding to the photometric error the same value of the amplitude variation estimated in the V band. We discuss about the implications of having wavelength-dependent inflated errors in Section~\ref{sec:caveats}.
Only a few of the carbon stars in the LMC were neither included in the catalogue by \citet{Riebel12} nor analysed in \citet{Groenewegen18}. These stars are fitted by employing the photometry contained in the catalogue by \citet{Jones17}.

The observed photometry is corrected for the interstellar reddening. A value of $A_{\rm V}=0.15$~mag \citep{Groenewegen18} and $A_{\rm V}=0.459$~mag \citep{Riebel12} is adopted for the SMC and for the LMC, respectively.
The assumed distances are $\sim 60$ \citep{Ngeow08} and $\sim 50$~kpc \citep{Cioni00, Keller06} for the SMC and the LMC, respectively.
\begin{table*}
\caption{References for the photometry adopted for the SED fitting of the carbon stars taken from different catalogues. The following acronyms hold: Magellanic Clouds Photometric Survey (MCPS);  Optical Gravitational Lensing Experiment (OGLE); Infrared Survey Facility (IRSF); Infrared Camera catalogue (IRC); Infrared Astronomical Satellite (IRAS); Wide-field Infrared Survey Explorer (WISE).}
\label{photometry}
\begin{center}
\begin{tabular}{l l l }
\hline
Carbon stars from &    Photometry &  References \\
\hline
\citet{Riebel12} &       MCPS & \citet{Zaritsky02, Zaritsky04} \\
                     &      2MASS                          & \citet{Skrutskie06}\\
                     &      Spitzer          &  \citet{Meixner06}    \\
\citet{Srinivasan16} &      MCPS & \citet{Zaritsky02, Zaritsky04} \\
                     &      OGLE                               &      \citet{Udalski08}\\
                     &      2MASS                          & \citet{Skrutskie06}\\
                     &      IRSF                           &  \citet{Kato07} \\
                     &      Spitzer          &  \citet{Meixner06, Boyer11, Boyer12}    \\
\citet{Jones17} &           MCPS & \citet{Zaritsky04} \\
                     &      2MASS                          & \citet{Skrutskie06}\\
                     &      IRSF                           &  \citet{Kato07} \\
                     &      Spitzer          &  \citet{Meixner06}    \\
\citet{Groenewegen18} & MCPS &  \citet{Zaritsky02, Zaritsky04}\\
                      & OGLE  & \citet{Udalski08a, Udalski08b}\\
                      & Bessel, Cousins & \citet{MASSey02}\\
                      & MACHO & \citet{Fraser08}\\
                      & Bessel, Cousins &  \citet{Wood83}  \\
                      & DENIS &  \citet{Cioni00, DENIS05}\\
                      & 2MASS, 2mass-6X & \citet{Skrutskie06};\citet{Cutri12}\\
                      & IRSF  &   \citet{Kato07, Macri15}  \\
                      & SAAO &  \citet{Whitelock89, Whitelock03} \\
                      & CASPIR & \citet{Wood98,Sloan06, Sloan08, Groenewegen07}\\
                      & IRAS        &  \citet{Moshir93}; \citet{Loup97}\\
                      & Spitzer & \citet{Meixner06, Bolatto07}\\
                      &         & \citet{Gruendl08, Whitney08, Gordon11}\\
                      & WISE & \citet{Wright10,Cutri13} \\
                      & Akari IRC   & \citet{Ishihara10, Ita10, Kato12}\\
\hline
\end{tabular}
\end{center}
\end{table*}
\section{SED fitting}\label{sec:fit}
For each of the models in the grid the reduced $\chi^2$ with respect to the observed fluxes is computed, similarly to what was done by \citet{Groenewegen09}, \citet{Gullieuszik12}, \citet{Riebel12}, and \citet{Srinivasan16}:

\begin{equation}\label{chi2}
\chi^2=\frac{1}{N_{\rm obs}}\sum_{i}\frac{(F_{i, {\rm obs}} - F_{i, {\rm th}})^2}{e_{i,{\rm obs}}^2},
\end{equation}
where $F_{i, {\rm th}}$ and $F_{i, {\rm obs}}$ are synthetic and observed fluxes for the $i$ band, $e_{i, {\rm obs}}$ is the associated error in each band and $N_{\rm obs}$ is the number of photometric points considered.

We note that the stars considered are all large amplitude variables and combination of photometry will thus always increase the $\chi^2$ of any model fit to the photometric data over that expected purely on the basis of the photometric errors.

The model that yields the best-fitting spectrum provides the stellar parameters of star considered. Some degeneracy in the  parameters is however present and not all the stellar quantities are well constrained by the SED fitting only \citep{Nanni18}. 
The uncertainty on each of the derived quantity is computed following \citet{Nanni18}. Briefly, the synthetic best-fitting photometric fluxes with $\chi^2=\chi^2_{\rm best}$ is randomly modified within the observed errors. The $\chi^2$ of the randomly-modified synthetic fluxes with respect to the observed ones is then recomputed. The same procedure is performed for $100$ sets of randomly-modified synthetic fluxes and $100$ $\chi^2$ values are obtained. From these $\chi^2$ the minimum and the 1-$\sigma$ values are extracted. The difference between these two values provides $\Delta\chi^2$ and $\chi^2_{\rm max}=\chi^2_{\rm best}+\Delta\chi^2$. All the synthetic spectra in the grids with $\chi^2\le\chi^2_{\rm max}$ represent the possible observed photometric fluxes. Therefore, the average value and the standard deviation $\sigma$ of every quantity is computed by including all the models in the grids with $\chi^2\le\chi^2_{\rm max}$. If the number of models that satisfy the condition $\chi^2\le\chi^2_{\rm max}$ is less than 4, we assume that the source is represented by the best-fitting value with zero uncertainty. 

Some of the synthetic spectra in the grids are excluded from the SED fitting procedure. In such models the assumed mass-loss rate is $\log\dot{M}\geq -5.5$ but the outflow is not accelerated via radiation pressure ($v_{\rm exp}=v_{\rm i}$). In these cases the mass loss assumed as input quantity is physically inconsistent with the fact that the outflow is not accelerated. The lower limit of the mass-loss rate of $\log\dot{M}\geq -5.5$ is taken from \citet{Andersen99}. In this work this is the minimum mass-loss rate obtained through hydrodynamic calculations with different optical data sets of amC dust. 
According to a recent investigation by \citet{McDonald2018}, a transition between pulsation enhanced and dust-driven is expected to occur at mass-loss rates above $\log\dot{M}=-6$.
Therefore the threshold limit for dust-driven wind selected in this work represents a safe assumption.
In case the wind is not accelerated for $\log\dot{M}<-5.5$ some mechanism different from dust-driven wind, i.e. magneto-acoustic and/or pulsation-driven wind, is assumed to produce the assumed mass-loss rate. For these models dust is passively condensed in the CSE that is moving at constant velocity.

Photometry with a relative error greater than 70\% is not taken into account in the SED fitting procedure.
The \textit{Spitzer} photometry at $4.5$ and $5.8$~$\mu$m is not included in the SED fitting calculations in the case where the star is only mildly dust-enshrouded ($\jks\lesssim 2$~mag). The spectra of these stars might be affected by the C$_3$ absorption features at those wavelengths \citep{Boyer11,Sloan15} that are not reproduced by the available opacity data sets \citep[see][Fig.~10]{Jorgensen00}. 
However, we only remove the $4.5$ and $5.8$~$\mu$m data if at least three photometric points to perform the SED fitting are left.

All photometry with filters centered at $\lambda> 20$~$\mu$m is excluded as well from the SED fitting, since the fluxes at these wavelengths can be affected by Magnesium Sulfide (MgS) emission \citep{Nanni16, Nanni18}, which is not included in our calculations. 
Indeed, despite the observational indication of MgS being a common species around carbon stars \citep{Zijlstra06, Sloan16}, its condensation process is difficult to explain on the theoretical point of view. \citet{Zhukovska08} have shown that a significant amount of MgS would be produced only in case this species can grow as a mantle on SiC grains or for extreme mass-loss rates,  $\dot{M}>5\times 10^{-4}$~M$_\odot$~yr$^{-1}$, if MgS is condensed as a separate dust species. 
However, in the former scenario the SiC feature around $11.3$~$\mu$m would be affected, while in the latter case the value of the mass-loss rate is much larger than the one estimated for the carbon stars in which the MgS feature is observed \citep{Sloan15}.
The exclusion of MgS from our calculations is not expected to affect the synthetic SED for $\lambda < 30$~$\mu$m.

Finally, photometry not associated with the TP-AGB (e.g. companion stars) is removed, as also discussed in Appendix A of \citet{Nanni18}. The photometry excluded for each star is available from the page \url{https://ambrananni085.wixsite.com/ambrananni/online-data-1}.
\section{Results}
\label{Results}
\subsection{Grids of models}\label{trends_tau}
\begin{figure}
\includegraphics[trim=0 0 0 0]{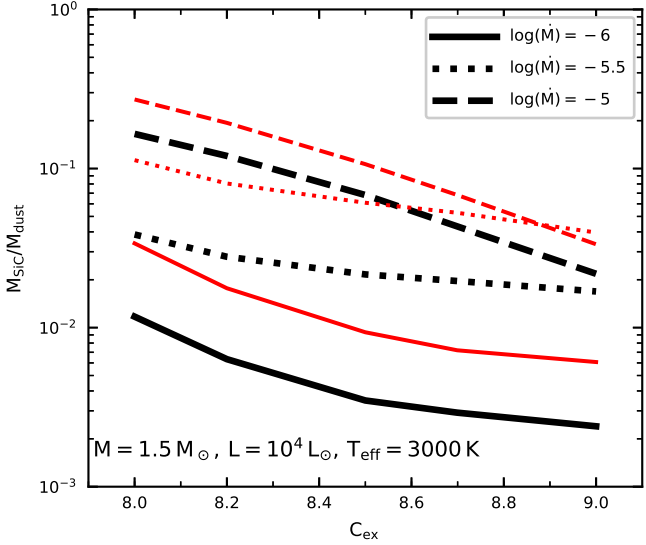}
        \caption{SiC mass fraction as a function of the carbon excess, for three choices of the mass-loss rate listed in the legend. Black thick lines represent the trends for $Z=0.004$, while the red, thinner, lines are computed for $Z=0.006$. All the models are computed for the H11 optical data set.}
        \label{sic}
        \end{figure}

Some results for the SMC and LMC grids of models are compared in this section.
In Fig.~\ref{sic} the SiC mass fraction as a function of the carbon excess is shown for different choices of the mass-loss rate.
The trends are very similar for the two different metallicity values.
The SiC mass fraction decreases with the carbon excess for both metallicity values because the amount of amC dust is increasing. On the other hand, for a given choice of the carbon excess, the SiC mass fraction increases with the mass-loss rate. Indeed, since SiC dust is
formed before the onset of the dust-driven wind and the subsequent density drop, larger initial densities favour a larger condensation fraction for this dust species.
For a given value of the carbon excess and of the mass-loss rate, the SiC mass fraction increases with the metallicity. This is not surprising since the metallicity determines the silicon abundance in the atmosphere.

\begin{figure}
        \includegraphics[trim=0 0 0 0]{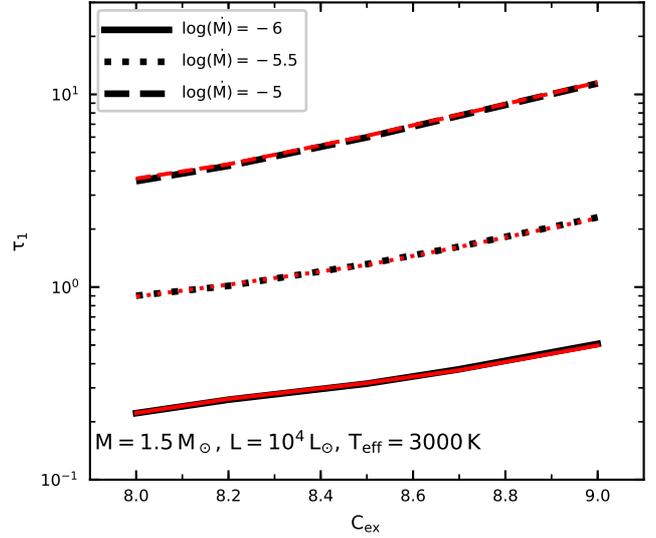}
        \caption{Optical depth at 1 $\mu$m ($\tau_1$) as a function of carbon excess and for different values of the mass-loss rate. The same metallicities of Fig. \ref{sic} are marked with the same line styles and colours. All the models are computed for the H11 optical data set.}
        \label{tau1_cex}
        \end{figure}

In Fig.~\ref{tau1_cex} the effect on the optical depth at 1~$\mu$m due to changing the metallicity values adopted is plotted as a function of $C_{\rm ex}$ for different mass-loss rates. The differences found between the two cases are negligible for the selected set of parameters.
The small difference found for the optical depth at 1~$\mu$m depends on the fact that the optical constants of the SiC are similar to the ones of the amC, except for the 11.3~$\mu$m feature \citep{Groenewegen98}.

\begin{figure}
\includegraphics[trim=0 0 0 0]{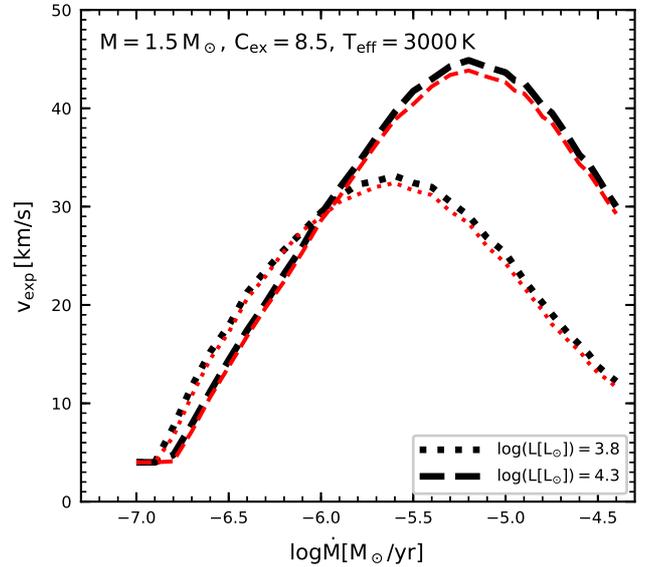}
        \caption{Expansion velocity as a function of mass-loss rate for different values of the luminosity listed in the legend. The same metallicity values of Fig.~\ref{sic} are marked with the same line styles and colours. All the models are computed for the H11 optical data set.}
        \label{vel_dmdt}
        \end{figure}

For the same reason, the trend between the expansion velocity and the mass-loss rate plotted in Fig.~\ref{vel_dmdt} is very similar, if only the metallicity value changes. This means that variations in the SiC content do not produce big modifications in the predicted CSEs.
The expansion velocity against the mass-loss rate in our grids of models shows a maximum and then declines. This trend is expected, since at increasing values of the mass-loss, the CSEs becomes progressively more optically thick, and the radiation pressure of photons on dust grains decreases \citep{Ivezic01, Ivezic10}. Since the optical depth decreases with the luminosity \citep[Fig.~2 in][]{Nanni18} the velocity peak is reached at larger mass-loss rates for higher luminosities.

From the results in Figs. \ref{tau1_cex} and \ref{vel_dmdt} it is possible to conclude that the optical depth around 1~$\mu$m and the dynamical properties of the outflow are not largely affected by differences of the adopted metallicity. 
This result is not surprising, since in C-stars the main properties of the outflow, except for the SiC abundance, are determined by the carbon-excess, rather than by the metallicity \citep{Mattsson10,Bladh19}.
For this reason, we also do not expect remarkable differences in our results if an $\alpha$-enhanced mixture of the elements or a more realistic metallicity distribution \citep{Nidever19} would be adopted in our calculations.
        
\subsection{Quality of the fit}
\begin{figure}
\includegraphics[scale=0.75]{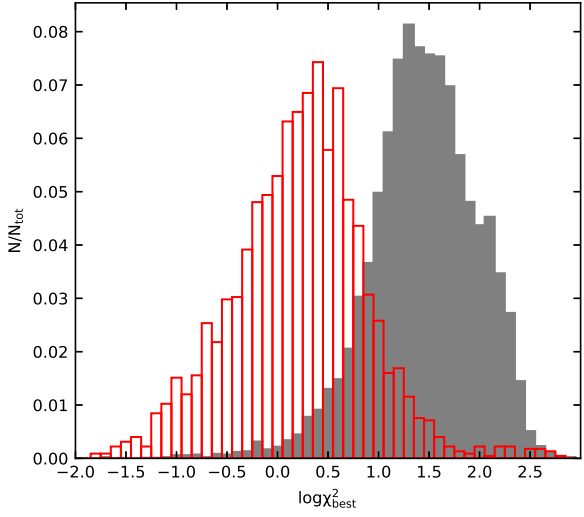}
        \caption{Normalized distributions of the $\chi^2_{\rm best}$ for the carbon stars in the LMC (grey) and in the SMC (red), plotted for the H11 optical data set.}
        \label{chi2_histo}
        \end{figure}

The $\chi^2_{\rm best}$ distributions for the carbon stars in the LMC and SMC are shown in Fig.~\ref{chi2_histo}.
The stars in the SMC are better fitted than the ones in the LMC.
The $\chi^2_{\rm best}$ is peaked around $\approx 3$ for the carbon stars in the SMC and between $20$ and $50$ for the stars in the LMC. The peak value of the $\chi^2_{\rm best}$ distribution for the stars in the LMC is similar to the one obtained by \citet{Riebel12}.
The different peak position of the stars in the SMC with respect to the LMC is related to the errors affecting the photometric fluxes of the stars in the two galaxies. Indeed, the errors on the fluxes are larger for the carbon stars in the SMC and consequently a value of the $\chi^2_{\rm best}$ lower than the ones in the LMC is usually obtained.

\begin{figure}
\includegraphics[trim=0 0 0 0, angle=90, width=0.48\textwidth]{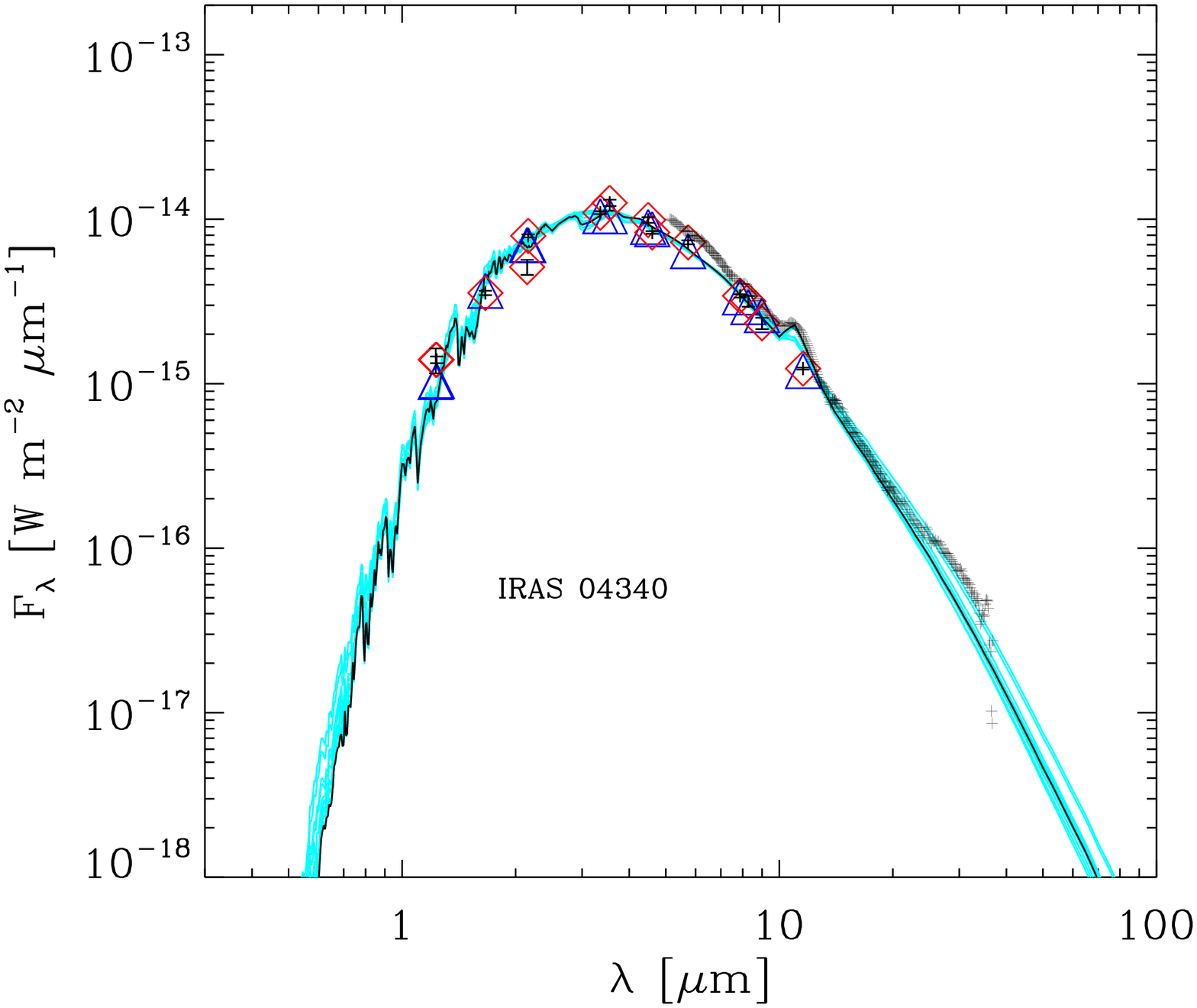}
\includegraphics[trim=0 0 0 0, angle=90, width=0.48\textwidth]{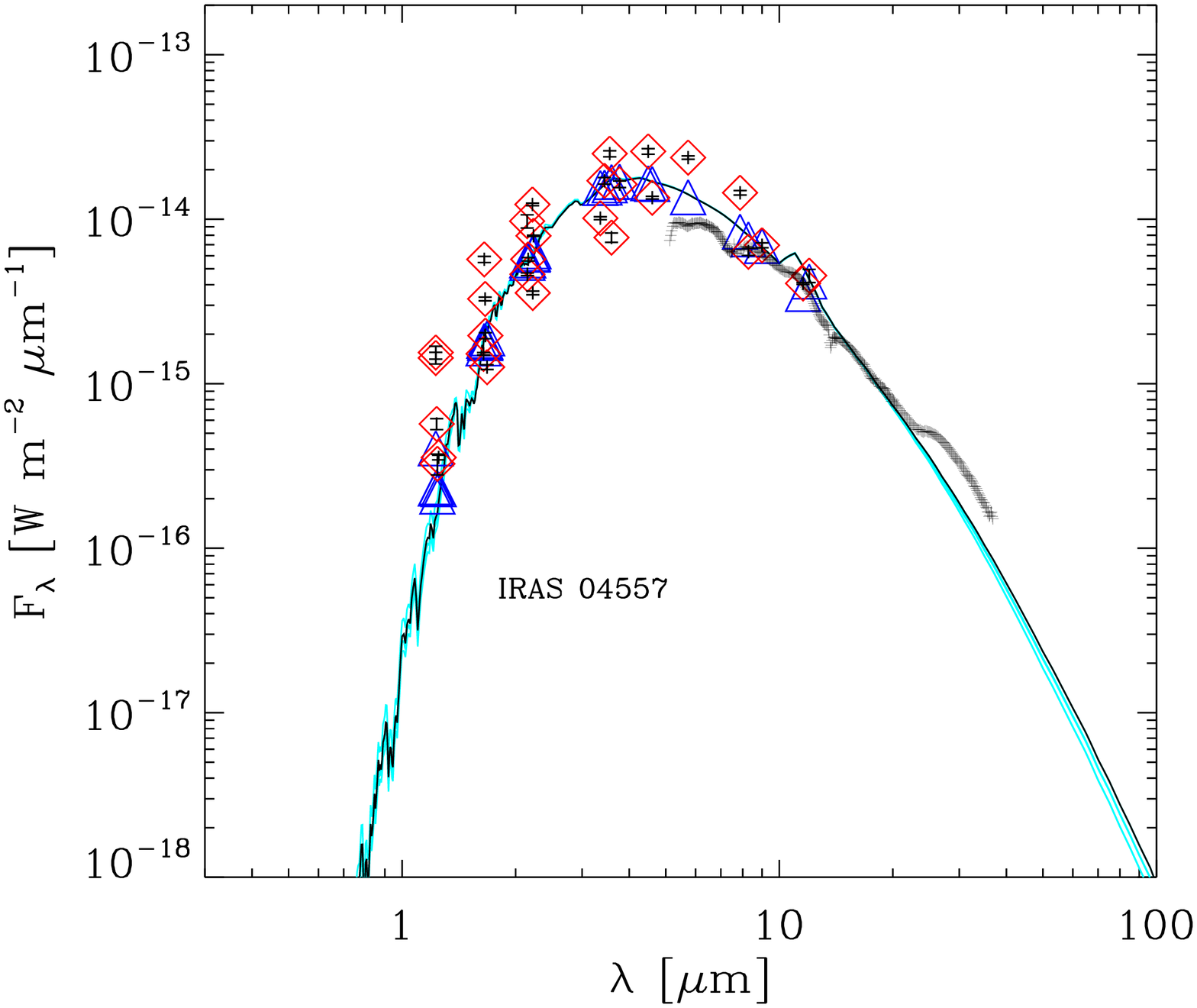}
        \caption{Examples of SED fitting for two dust-enshrouded stars. The observed photometry and the associated errors are marked with red diamonds and black bars, respectively. The IRS spectra are overplotted with black crosses. The synthetic photometry is plotted with blue triangles, the best-fitting spectrum is shown in solid black while the cyan lines are the acceptable spectra (see text for more details).}
        \label{fit_spectra}
        \end{figure}

The sources from the catalogue by \citet{Groenewegen18} are fitted by employing all the photometry listed in Table~\ref{photometry} and they are typically characterized by  $\chi^2_{\rm best}$ larger than the other sources. The photometric fluxes of these stars are not observed at the same epoch and they can be different for similar wavelengths.
In Fig.~\ref{fit_spectra} two examples of the fitted carbon stars in the LMC from \citet{Groenewegen18} are shown. The observed fluxes of IRAS F04340--7016 (IRAS 04340), upper panel, are much less scattered than the ones of IRAS 04557--6753 (IRAS 04557), lower panel, and this is reflected in the different values of the $\chi^2_{\rm best}$ for these two stars. Fig. \ref{fit_spectra} illustrates how the fit obtained is in reasonable agreement with the observations, but for IRAS 04340 $\chi^2_{\rm best}\sim 10$, while for IRAS 04557 $\chi^2_{\rm best}\sim 185$.
\subsection{Stellar and dust properties}
From the SED fitting procedure it is possible to derive the stellar and dust properties that are discussed in the following.
\subsubsection{Luminosity function}\label{sec:luminosity_f}
\begin{figure}
\includegraphics[scale=0.75]{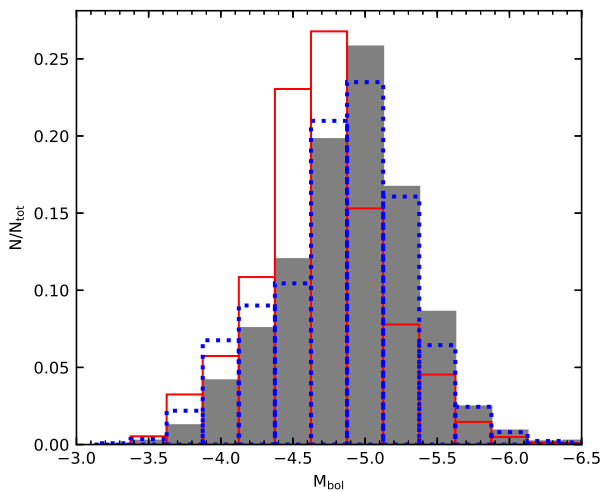}
        \caption{Luminosity function of carbon stars
derived from the SED fitting procedure for the LMC (grey histogram) and for the SMC (solid red line). The dotted blue line is the luminosity function derived by \citet{Riebel12}.}
        \label{L_func}
        \end{figure}

In Fig.~\ref{L_func} the normalised luminosity function of the entire sample of carbon stars is shown for the MCs.
Our luminosity function for the LMC sources is in very good agreement with the one derived by \citet{Riebel12}.
An excellent agreement was also found between the luminosity function derived by \citet{Srinivasan16} and the one computed in \citet{Nanni18} for the SMC. Furthermore, \citet{Riebel12} and \citet{Srinivasan16} derived the luminosities adopting the same distances as the ones assumed in this work.
The luminosity functions of the SMC and LMC are similar in shape, however the carbon stars in the LMC are shifted to higher luminosity, $-6\lesssim$M$_{\rm bol}\lesssim -3.5$ mag, with respect to the ones in the SMC. For the SMC M$_{\rm bol}$ peaks around $\approx-4.75$ mag while the peak for the LMC is M$_{\rm bol}\approx -5$ mag. 
As already noticed in \citet{Srinivasan16}, the shift in the luminosity peak can be due to a larger mass range for carbon stars at low metallicity combined with the effect of different star formation histories in the two galaxies.
\subsubsection{Mass-loss rates}\label{sec:ml}
\begin{figure}
\includegraphics{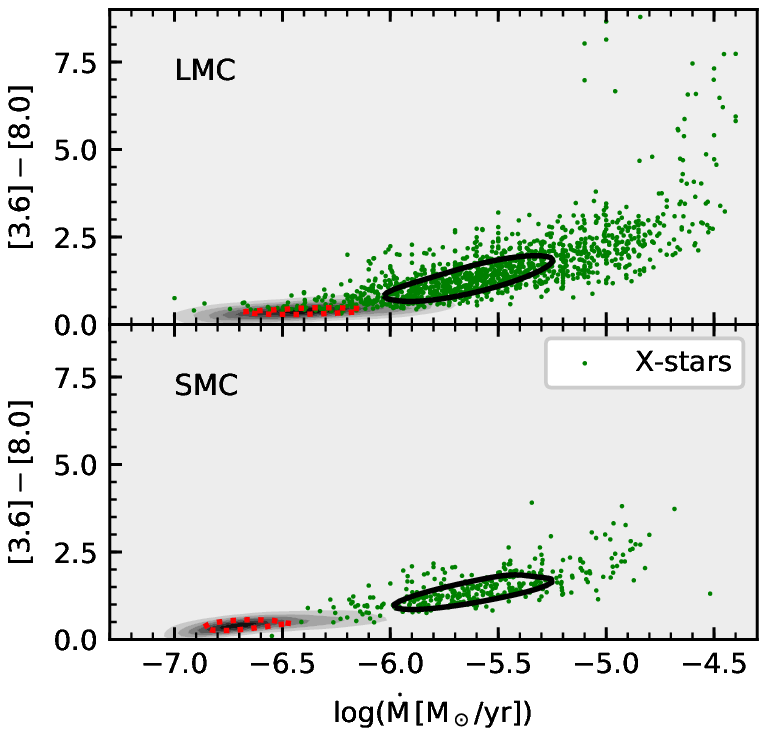}
\includegraphics{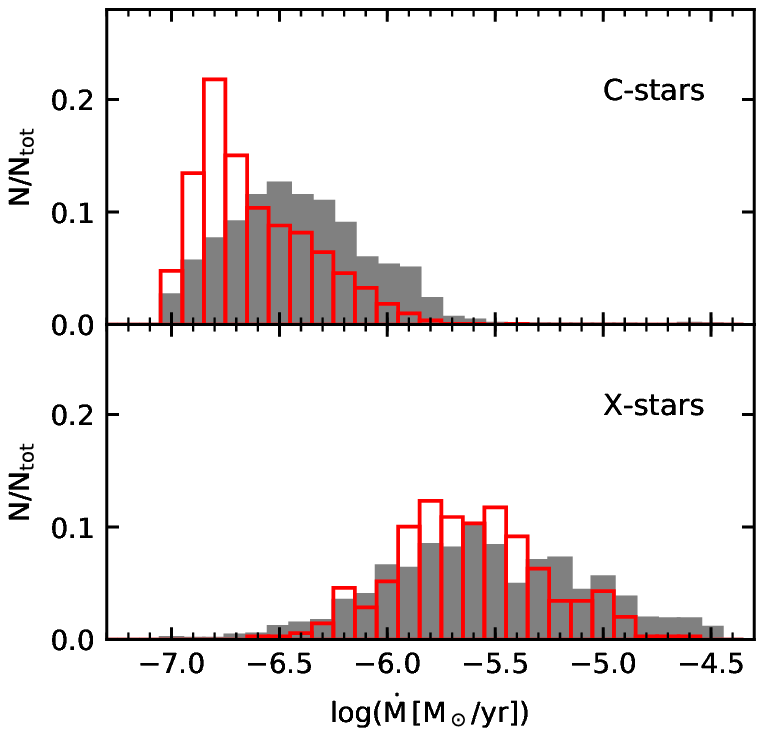}
        \caption{Upper panel: observed $[3.6]-[8.0]$ colour as a function of the mass-loss rate for the carbon stars in the MCs. The linear normalized density map from 0, light grey, to 1, black, includes all the carbon stars. X-stars are over plotted with green symbols.
The stars photometrically classified as C and X are contour plotted with dotted red and solid black lines, respectively. Lower panel: The normalized distribution of mass-loss rates for the stars in the LMC (grey histogram) and SMC (solid-red line).}
        \label{mloss_histo}
        \end{figure}

In Fig.~\ref{mloss_histo} the observed $[3.6]-[8.0]$ colour as a function of the mass-loss rate is shown together with the corresponding normalised distributions for C- and X-stars. This colour is selected since usually also the most dust-enshrouded sources are detected in the $[3.6]$ and $[8.0]$ bands. The case for the H11 optical data set is shown, but the J1000 provides similar trends. 
The results for the LMC and for the SMC are compared.
The separation between the stars classified as C or X occurs around $\log{\dot{M}}\approx -6$, in agreement with our previous analysis \citep{Nanni18}.
From the upper panel of Fig.~\ref{mloss_histo} it is possible to see how a large variation in the selected colour (between $2.5$ and $8$ mag) occurs in a narrow range of the mass-loss rate $-4.7\lesssim\log{\dot{M}}\lesssim -4.4$ for some of the X-stars in the LMC. Such extreme colours are not observed for the X-stars in the SMC. About one-third of the X-stars with such a mass-loss rate are spectroscopically classified as carbon-rich by \citet{Jones17} and \citet{Groenewegen18}. The classifications are consistent with the exception of only one source, classified as Young Stellar Object by \citet{Jones17}.
Some of the sources which lack a spectroscopic classification might be OH/IR stars that are present in the LMC but that are missing in the SMC \citep{Goldman18}.
As can be also seen from the lower panel of Fig.~\ref{mloss_histo}, the normalised distribution of the X-stars in the LMC exhibits an excess of stars with mass-loss rates larger than $\log{\dot{M}}\sim -5.3$ with respect to the SMC. In particular, a non-negligible fraction of X-stars in the LMC have mass-loss rates larger than their counterparts in the SMC. 
Also for the C-stars larger mass-loss rates are predicted for the sources in the LMC, while the distribution of these stars in the SMC peaks towards lower values.
The difference in the predicted mass-loss rates for the two galaxies can be due to several factors. One possible explanation is that carbon stars in the SMC might be characterised by lower mass-loss rates because of their lower luminosity, as presented in Fig.~\ref{L_func}. Star formation history may also play a role.
\subsubsection{Gas-to-dust ratios}\label{sec:dtg}
\begin{figure}
\includegraphics{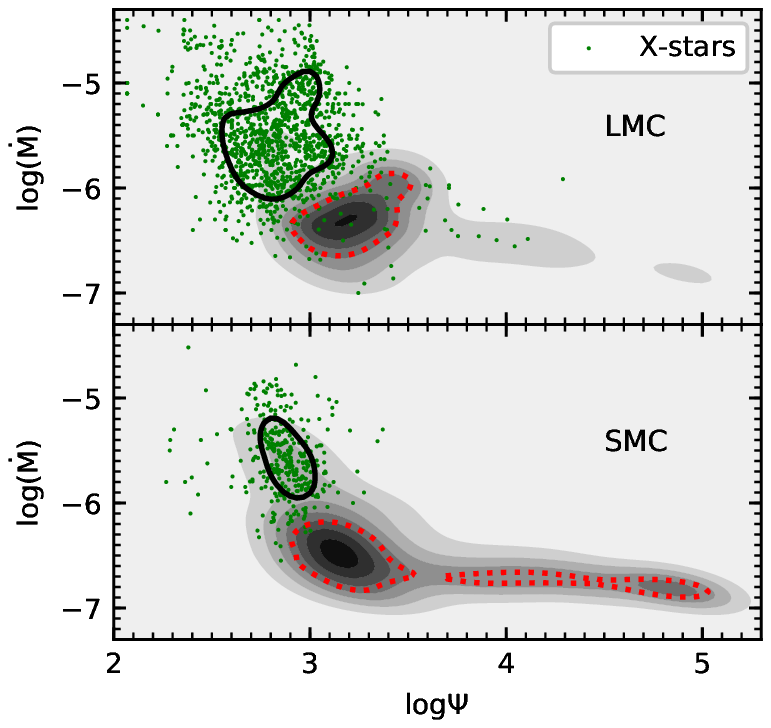}
\includegraphics{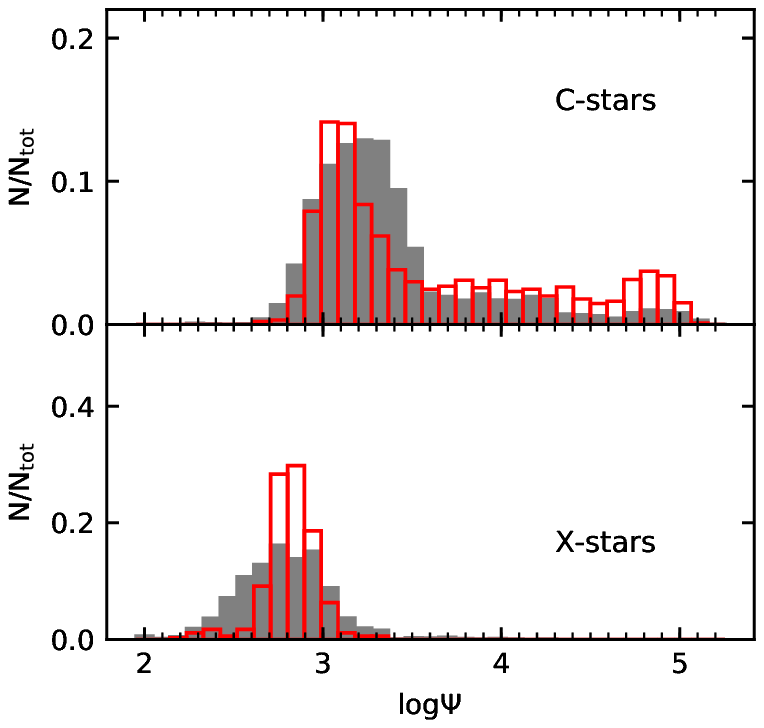}
        \caption{Upper panel: the mass-loss rate as a function of the gas-to-dust ratio for the carbon stars in the MCs, computed with the H11 data set. The same symbols and line style as for Fig.~\ref{mloss_histo} are adopted. Lower panel: the normalized distribution of the gas-to-dust ratio for the stars in the LMC (grey histogram) and SMC (solid-red line).}
        \label{gtd_histo}
        \end{figure}

In Fig.~\ref{gtd_histo} the gas-to-dust ratio ($\Psi$) as a function of the mass-loss rate is shown for the two galaxies, together with their normalised distributions. The case for the H11 data set is shown.
From the upper panel of Fig.~\ref{gtd_histo} it is possible to notice that the gas-to-dust ratio covers a large range of values with similar trends for the two galaxies, from almost dust-free for the lowest mass-loss rates, to heavily dust-enshrouded for increasing mass-loss rates. 
The value of $\Psi$ decreases with the mass-loss rate in agreement with our previous analysis of the SMC sources \citep{Nanni18}, and with hydrodynamic simulations in a similar range of carbon-excess and mass-loss rates \citep{Mattsson10, Eriksson14}. The value of $\Psi$ flattens around $\log\dot{M}=-6$, where the stars become extreme and more dust-enshrouded.
The X-stars in the LMC show a spread in the values of the gas-to-dust ratios larger than the ones in the SMC for a given value of the mass-loss rate. For both galaxies the typical value of the gas-to-dust ratio for X-stars is around $\sim 700$, as can be also seen from the lower panel of the same figure. With respect to \citet{Nanni18} the value of gas-to-dust is lower because of the larger carbon-excess. 

A gas-to-dust ratio of $\sim 700$ is larger than the one assumed in the literature ($200$) by different authors \citep{Groenewegen09, Boyer12, Matsuura13}. In other works the gas-to-dust is of $500$ and $1000$ in the LMC and the SMC, respectively, since this value is assumed to vary with the metallicity of the galaxy \citep{vanLoon99, vanLoon05b}. 
The typical value we find in our analysis for the extreme carbon stars is thus closer to the ones adopted by \citet{vanLoon99} and \citet{vanLoon05b}.
Our finding implies that the predicted condensation efficiency is typically $\sim 3$ times smaller than the one assumed if $\Psi=200$.
For the most mass-losing stars in the LMC, the gas-to-dust ratio is sometimes lower than the one derived for the SMC for the same mass-loss rate. This implies that a higher efficiency in the dust condensation is expected to occur for some of the extreme sources in the LMC with no counterparts in the SMC. 
This finding is in agreement with the trends derived by \citet{vanLoon00}. In another investigation, \citet{vanLoon08} interpreted similar acetylene absorption features in the SMC and LMC carbon stars as the result of comparable abundances in the gas phase in their CSEs. However, at similar acetylene features corresponded stars redder in the LMC than in the SMC. The interpretation was that amC dust condenses less efficiently in CSEs of carbon stars in the SMC.

In the analysis presented here, the lowest gas-to-dust attained in the LMC is $\sim100$ for $-5 \lesssim\log{\dot{M}}\lesssim -4.4$, a value even lower than the common assumption found in the literature ($200$). On the other hand, the gas-to-dust ratio for the X-stars in the SMC is down to $\sim 160$--$200$ for a few stars with $-6 \lesssim\log{\dot{M}}\lesssim -4.5$. This lower limit of the gas-to-dust ratio is smaller than the ones derived in \citet{Nanni18}, but such small values are predicted only for a few stars. The difference with respect to our previous analysis is not surprising since in the present work larger values of the carbon-excess have been included in the grids. 

The gas-to-dust ratio distributions of the C-stars show a different behaviour with respect to the X- sources in the two galaxies. For the C-stars in the SMC the distribution is shifted towards lower values with respect to the one derived for the LMC, while the opposite holds for X-stars. The distribution  is also tighter for the C-stars in the SMC. For C-stars the result seems to be dependent on the choices of the optical data set, since for the J1000 the distributions are similar in the two galaxies.
\subsubsection{Dust chemistry}
\begin{figure}
\includegraphics{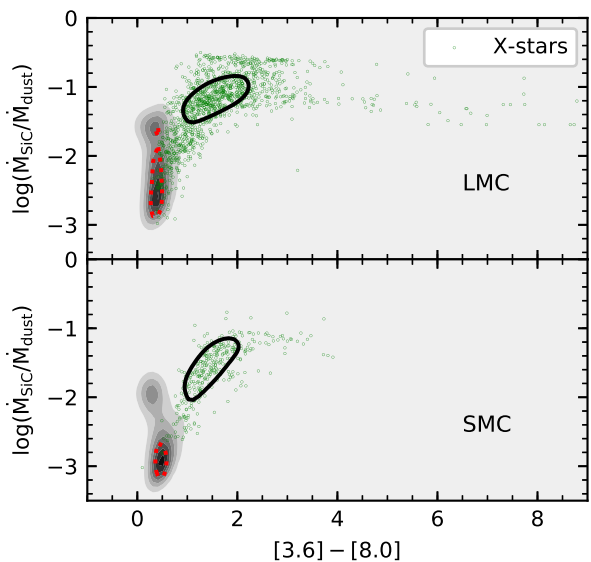}
        \caption{The SiC mass fractions as a function of the observed $[3.6]-[8.0]$ colour derived for the H11 optical data set for the LMC and the SMC. The same symbols and line style as for Fig.~\ref{mloss_histo} are adopted.}
        \label{dust_chem}
        \end{figure}

In Fig.~\ref{dust_chem} the SiC mass fraction is shown as a function of the observed $[3.6]-[8.0]$ colour. The case with the H11 set is shown, since the results are similar for the J1000 one. The SiC mass fraction is usually less than one per cent for C-stars, while it linearly increases for X-stars with a plateau around $[3.6]-[8.0]\approx 2$~mag where it reaches a maximum fraction of $\sim 30$ per cent, for the LMC, and of $\sim 10$ per cent, for the SMC. This latter value is in agreement with our previous analysis for which the value of $C_{\rm ex}$ was limited to $8.5$ \citep{Nanni18}. 
The larger SiC fraction obtained for the X-stars in the LMC is due to the larger metallicity adopted in the calculations, as also shown in Fig.~\ref{sic}. This result is qualitatively in good agreement with different observations \citep{Sloan06,Zijlstra06, Lagadec07, Leisenring08, Sloan15}, where the increase in the strength of the SiC feature with respect to the continuum emission in galaxies with higher metallicity, is interpreted as an augmentation in the SiC content with metallicity. 
\subsubsection{Expansion velocities}\label{V_scaling}
\begin{figure}
\includegraphics{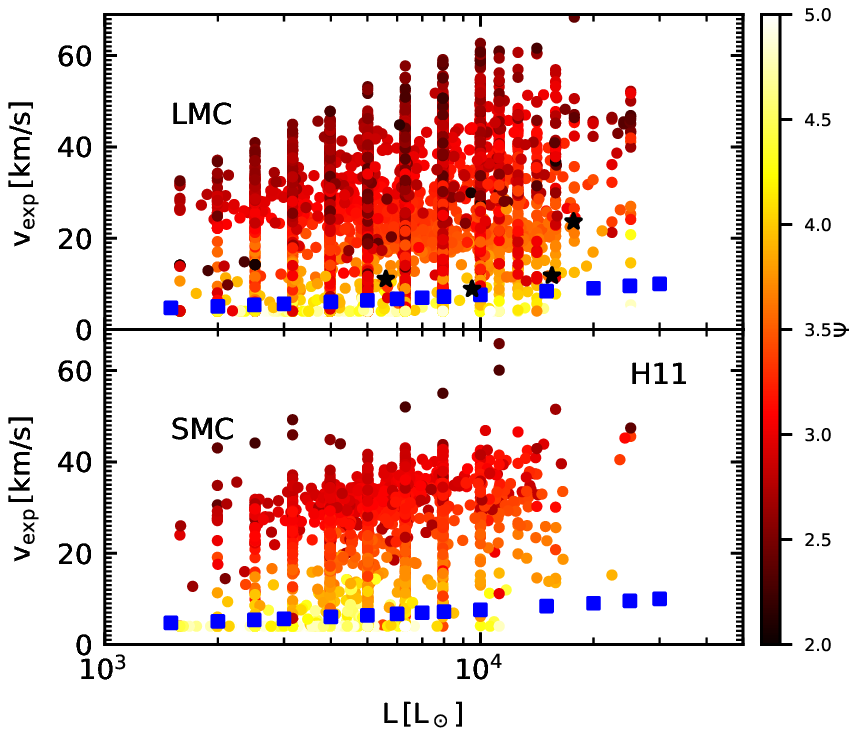}
\includegraphics{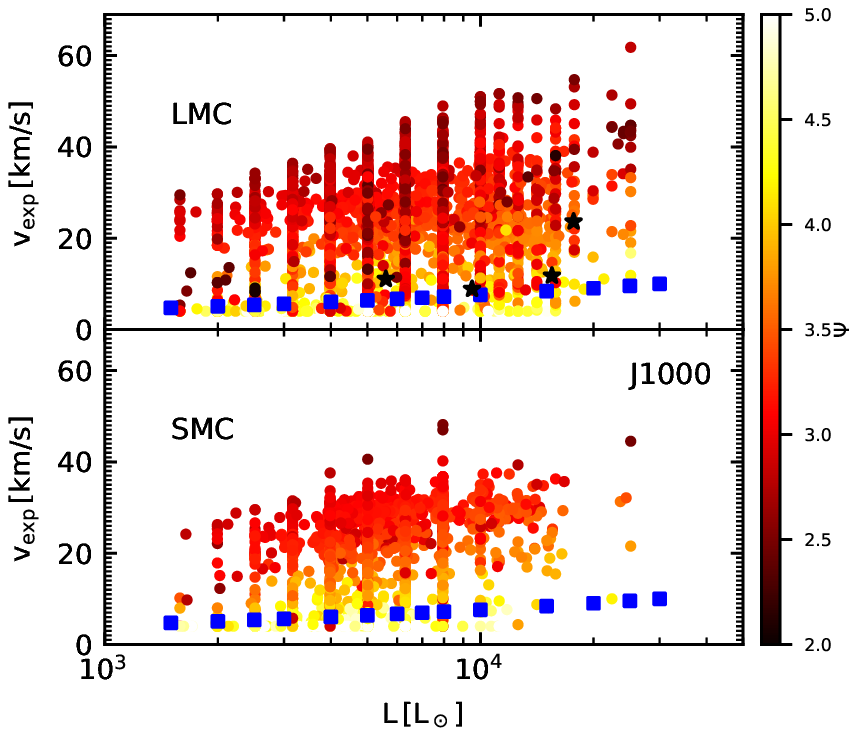}
        \caption{The outflow expansion velocity as a function of the luminosity derived for the H11 (upper panel) and for the J1000 (lower panel) optical data sets for the carbon stars in the LMC and in the SMC. Stars are colour-coded according to their gas-to-dust ratio. Blue squares indicate the assumed expansion velocity in \citet{Srinivasan16} that scales with the luminosity and gas-to-dust ratio as in equation~\ref{vexp_scaled}. Black star-like symbols represent the carbon stars observed in the LMC by ALMA \citep{Groenewegen_etal16}.}
        \label{vel_l_MCs}
        \end{figure}

\begin{figure}
\includegraphics{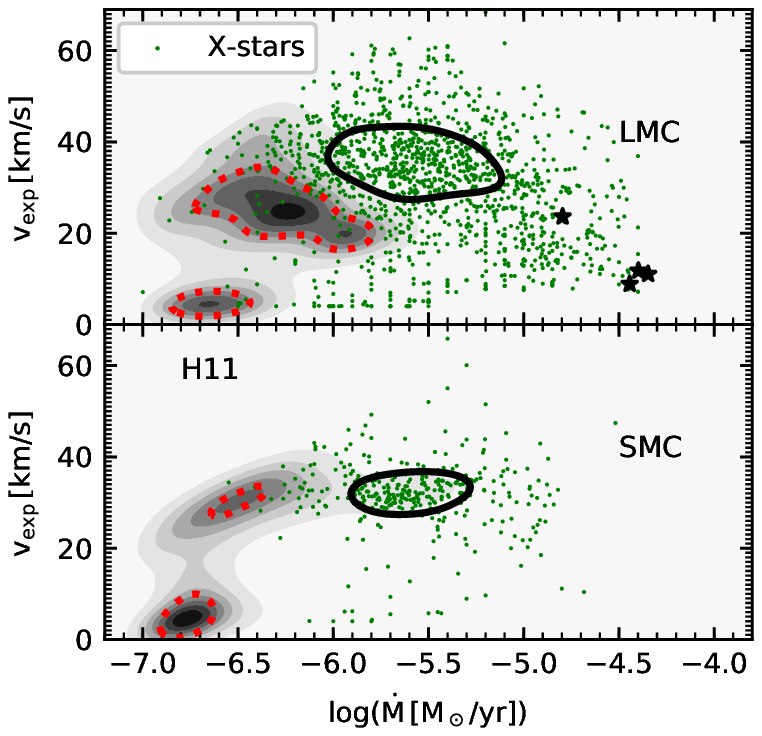}
\includegraphics{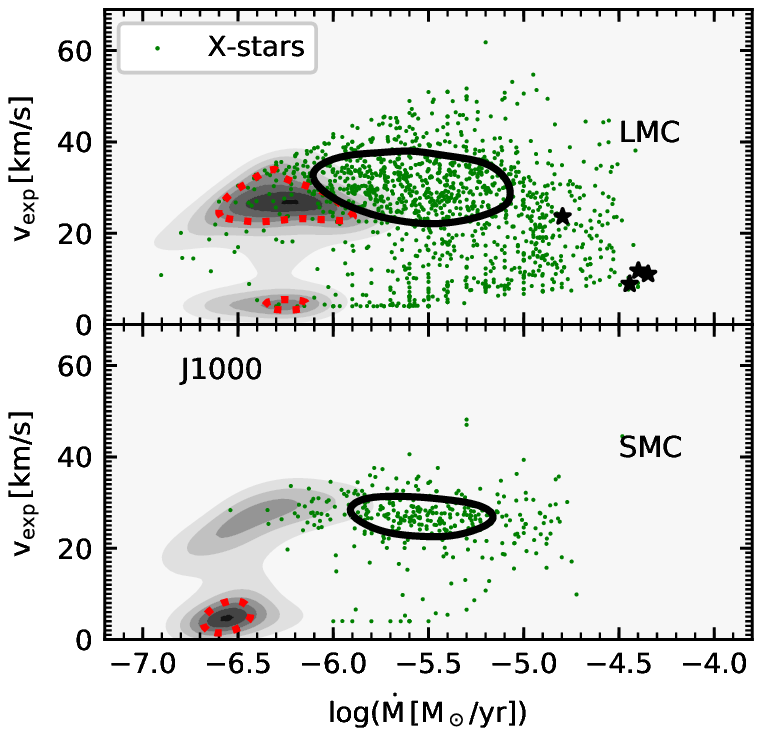}
        \caption{The outflow expansion velocity as a function of the mass-loss rate derived for the H11 (upper panel) and for the J1000 (lower panel) optical data sets for the carbon stars in the LMC and in the SMC. The same symbols and line style as for Fig.~\ref{mloss_histo} are adopted. Black star-like symbols represent the carbon stars observed in the LMC by ALMA \citep{Groenewegen_etal16}.}
        \label{vel_dmdt_MCs}
        \end{figure}

In Figs.~\ref{vel_l_MCs} and \ref{vel_dmdt_MCs} the maximum outflow expansion velocity achieved in the CSE as a function of the luminosity and of the mass-loss rate respectively are shown for the two selected optical data sets. Blue squares indicate the assumed expansion velocity in \citet{Srinivasan16} that scales with the luminosity and with the gas-to-dust ratio as in \citet{vanLoon06_2}:
\begin{equation}\label{vexp_scaled}
v_{\rm exp}\propto\left(\frac{L}{L_\odot}\right)^{1/4} \left(\frac{\Psi}{200}\right)^{-1/2},
\end{equation}
where $v_{\rm exp}=10$~km~s$^{-1}$ for a star with luminosity $L= 30000$~$L_\odot$ and $\Psi$ is the gas-to-dust ratio assumed to be $200$ for all the sources. 
The four carbon stars in the LMC for which the expansion velocities are derived from CO line observations performed with the Atacama Large Millimeter Array (ALMA) by \citet{Groenewegen16} are indicated with star-like symbols. The luminosities and mass-loss rates of these stars are derived from the SED fitting performed in this work. 
For both galaxies the expansion velocity shows a trend with the luminosity and the gas-to-dust ratio, with larger wind speed to more dust-enshrouded stars, even though the spread of values is large. 
In both galaxies a considerable fraction of stars shows low wind speeds ($<10$~km~s$^{-1}$) for both the optical data sets. Most of these sources are photometrically classified as C-stars and are characterized by large values of the gas-to-dust ratio ($\log\Psi>4$).
The outflows of these stars are either not accelerated via a dust-driven wind or only mildly accelerated. The wind speeds for these stars appear to be comparable with the relation adopted by \citet{Srinivasan16}. A fraction of C-stars and almost all X-stars are instead deviating from both \citet{Srinivasan16} and from \citet{Groenewegen06}'s assumption of constant outflow velocity of $\sim 10$~km~s$^{-1}$. These stars are more dust-enshrouded and attain expansion velocities $\sim 30$~km~s$^{-1}$ already at $L\sim 2000$--$3000$~L$_\odot$.
For the same value of the luminosity, the carbon stars in the LMC reach values of $\Psi$ lower than the ones of the SMC sources. For this reason, the stars in the LMC attain a maximum value of the velocity larger than the ones in the SMC. In particular, the faster wind speeds are reached for the stars with the lowest gas-to-dust ratio.

The trends derived are similar for the two optical data sets, but larger velocities are obtained for the H11 set. For this data set the wind speeds attained are up to $\sim 60$~km~s$^{-1}$ at $L\sim10000$~L$_\odot$ for the LMC, while for the same luminosity $v_{\rm exp}$ is $\sim 50$~km~s$^{-1}$ for the J1000 set. For luminosities larger than $L\sim10000$~L$_\odot$, the maximum value of the velocity of X-stars in the LMC tends to slightly decrease for the H11 data set, while it remains approximately constant for the J1000. This trend is less obvious for the X-stars in the SMC.

The stars in the LMC and SMC show similar trends in the $v_{\rm exp}$ against mass-loss rate plot shown in the two panels of Fig.~\ref{vel_dmdt_MCs}. The expansion velocity increases with the mass-loss rate until a maximum value is reached around $-5.5\lesssim\log{\dot{M}}\lesssim -5$. The typical value attained for the X-stars in the SMC is $\sim30$~km~s$^{-1}$ for both data sets, while the X-stars in the LMC are characterised by velocities between $\sim 40$~km~s$^{-1}$, for the H11 data set, and $\sim30$~km~s$^{-1}$, for the J1000.
After reaching its maximum value, the velocity declines, reflecting the behaviour in of the models in the grids discussed in Section~\ref{trends_tau}.
A behaviour similar to the one predicted by our analysis and by other authors \citep{Ivezic01, Ivezic10} can be pinpointed also for the carbon stars observed in our Galaxy \citep[see][Figs.~16 and 18]{Nanni18}. Specifically, in the sample by \citet{Groenewegen02} shown in the aforementioned figures, few stars around $\log\dot{M}=-5$ and $\log L=3.8$ seem to attain a maximum of the wind speed, even though the value reached is lower than in our analysis. 
The locations of the peak in $\log L$ and $\log\dot{M}$ is in reasonable agreement with the ones we found for the LMC sources. On the other hand, our results appear to be at odd with the wind speed derived for other samples of Galactic carbon stars \citep{Olofsson93, Schoier01,RamstedtOlofsson_14, Danilovich15}. This discrepancy might depend on the properties of the selected sample. For example, stars with larger metallicity might be characterised by lower values of carbon-excess that can affect the final expansion velocity, as well as the mass-loss and luminosity at which the star becomes optically thick. 

A fraction of C-stars with $\log\dot{M}<-6.4$ (SMC) and $\log\dot{M}<-6.2$ (LMC), for the H11 set, and $\log\dot{M}<-6.2$ (SMC) and $\log\dot{M}< -6$ (LMC) for the J1000, exhibit expansion velocities $<10$~km~s$^{-1}$.
The velocities predicted for the SMC for the two sets of optical constants are larger than in our previous analysis \citep{Nanni18}, because of the larger value of the carbon-excess adopted in these grids of models.

For a fraction of carbon stars in the MCs the predicted expansion velocity can be significantly larger, up to $60$~km~s$^{-1}$, than the one observed for Galactic carbon stars that rarely exceed $35$~km~s$^{-1}$ \citep[see][Fig.~18]{Nanni18}. Such large values of the wind speed in the MCs should be confirmed by direct observations. If confirmed, the difference between the wind speed of carbon stars in the MCs and in the Milky Way can be ascribed to their different carbon-excess. Indeed, in our Galaxy, carbon stars are expected to attain lower values of carbon-excess due to their larger metallicity.
For the X-stars observed by ALMA, for which the wind speed has been derived from CO line observations, the expansion velocities are reproduced by our approach.
The aforementioned stars will be studied in detail in Section \ref{subsc:alma}.
\subsection{Mass-loss rates and DPRs of individual stars: comparison with the literature}\label{literature}
\begin{figure}
        \includegraphics{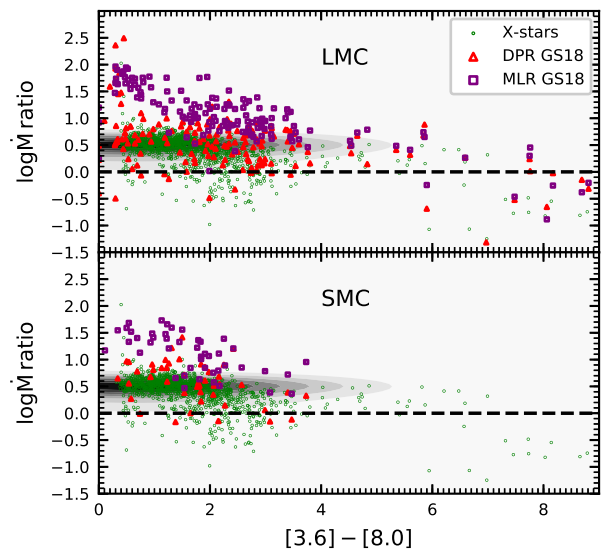}
        \caption{Ratios between the DPR or mass-loss rates (MLR in the figure) computed for the H11 data set and the same quantities derived by other authors, as a function of the $[3.6]$--$[8.0]$ colour. The DPRs for the LMC and the SMC are from \citet{Riebel12} and \citet{Srinivasan16}, respectively. The DPRs and mass-loss rates from \citet{Groenewegen18} are also shown:  purple squares are the ratio of the DPRs, while red triangles represent the ratio for the total mass-loss rates.}
        \label{mloss_histo_gro18}
        \end{figure}

In Fig.~\ref{mloss_histo_gro18} the ratios between our DPRs and/or mass-loss rates and the same quantities computed by \citet{Groenewegen18,Riebel12, Srinivasan16} are shown for the H11 data set. 
Similar trends are obtained for the J1000 set.
For consistency, we exclude from the analysis those stars classified as oxygen-rich by \citet{Riebel12} and \citet{Srinivasan16}.
For most of the stars our DPR is typically $\sim3$ times larger than the DPR derived by \citet{Riebel12} and \citet{Srinivasan16}.
These works are based on the GRAMS grids by \citet{Srinivasan11} in which the optical constants for amC dust measured by \citet{Zubko96} are employed. The optical constants adopted have been measured for carbon grains produced by an arc discharge between amorphous carbon electrodes in an Ar atmosphere at 10 mbar (ACAR sample).
For the estimate of the DPRs the wind speed is assumed to be of $10$~km~s$^{-1}$, in \citet{Riebel12} and scaled with the luminosity as in equation~\ref{vexp_scaled} for \citet{Srinivasan16}. The dust-density profile is assumed to be $\propto r^{-2}$.
The ratios between our DPRs and the DPRs by \citet{Riebel12} and \citet{Srinivasan16} show a linear trend with the $[3.6]-[8.0]$ colour that reflects the trend between the ratio of our predicted wind speed and the one assumed in the aforementioned works. The final result also depends on the optical data sets and on the different dust-density profiles adopted. In case the wind speed derived in our approach is overestimated, the differences between our DPRs and the ones by \citet{Riebel12} and \citet{Srinivasan16} would be reduced.
Specifically, in our calculations the expansion velocities, gas-to-dust ratio and dust-density profiles are consistently computed.
The luminosity distributions obtained in this work are instead in very good agreement with \citet{Riebel12} and \citet{Srinivasan16}, as discussed in Section~\ref{sec:luminosity_f}.
For a few X-stars in the LMC our DPR is down to $\sim10$ times smaller than the DPR by \citet{Riebel12} for the most dust obscured sources.
This difference can be explained if slow winds for these stars are predicted by our analysis. 
However, for only six stars the predicted expansion velocity is lower than $10$~km~s$^{-1}$, with a minimum value of $5$~km~s$^{-1}$ for two stars. Such a value of the wind speed would yield in our analysis half of the DPR derived by \citet{Riebel12} for the same stars. The difference in the expansion velocities is thus not sufficient to explain the result.
We notice that for these sources the $\chi^2_{\rm best}$ is usually high and the best fit is obtained for the largest values of optical depth in our grids, thus the DPR of these sources might have been underestimated.

A comparison of our mass-loss rates and the estimates by \citet{Riebel12} and \citet{Srinivasan16} would be possible only by assuming a fixed value of the gas-to-dust ratio, since only the DPRs are listed in their catalogues.

In \citet{Groenewegen18} the optical constants are computed for a continuous distribution of hollow spheres to take into account the possible porosity of dust grains, and are based on the optical constants measured by \citet[][ACAR sample]{Zubko96}. 
The DPRs in \citet{Groenewegen18} are estimated by assuming a constant expansion velocity of the outflow of $10$~km~s$^{-1}$.
The mass-loss rates are then derived from the DPRs by assuming a single value of the gas-to-dust ratio of $200$ for all the stars.
Our DPRs are larger than the DPRs from \citet{Groenewegen18} for more than $\sim90$ per cent of the carbon stars in the MCs. 
Furthermore, our estimate of the mass-loss rates is larger than the one obtained by \citet{Groenewegen18} for $\approx 95$ per cent of the stars in the LMC and for all the sources in the SMC. Also in this case, we find a trend between the DPR ratios and the colour.
The trend between the mass-loss rate and the colour is even more evident, and is due to the combined behaviour of the wind speed and of the gas-to-dust ratio. In our analysis, these quantities evolve with the stellar parameters, while they are constant in \citet{Groenewegen18}. 
For few stars among the most dust enshrouded, our DPRs and mass-loss rates are up to $\sim30$ times smaller than in \citet{Groenewegen18}. Also in this case, these stars are fitted by spectra corresponding to the largest optical depth of our grids. Consequently their DPRs and mass-loss rates might have been underestimated by our analysis. Again, low values of wind speed cannot explain alone the difference found.
\subsection{Total dust production rates}\label{tot_DPR}
The total DPRs and the associated uncertainties computed as described in Section~\ref{sec:fit} are listed in Table~\ref{DPR} together with the DPRs from the literature.
The total DPRs for the SMC are about $1.3$ times larger than the ones from our previous work \citep{Nanni18}, but still in agreement within the uncertainties. In our analysis, X-stars constitute more than $80$ per cent of the total budget of carbon stars in the MCs.
The SiC total mass fraction is between $\sim 3$ and $4$ per cent for the SMC and $\sim 8$ per cent for the LMC.
The iron mass fraction is always less than one per cent. The low amount of iron dust can be explained by the condensation temperature of the different dust species around carbon stars. SiC is the first dust species that forms, followed by carbon dust that is the driver of the outflow acceleration.
Iron dust is condensed after the onset of the dust-driven wind and the drop of the density that suppresses further dust condensation.
The DPRs computed with the two optical data sets are comparable within the uncertainties.
The representative value of the total DPR is derived by averaging the DPRs of the two data sets. The values obtained are $\sim1.8\times10^{-5}$~M$_{\odot}$~yr$^{-1}$, for the LMC, and $\sim2.5\times10^{-6}$~M$_{\odot}$~yr$^{-1}$, for the SMC.

\begin{figure}
\includegraphics{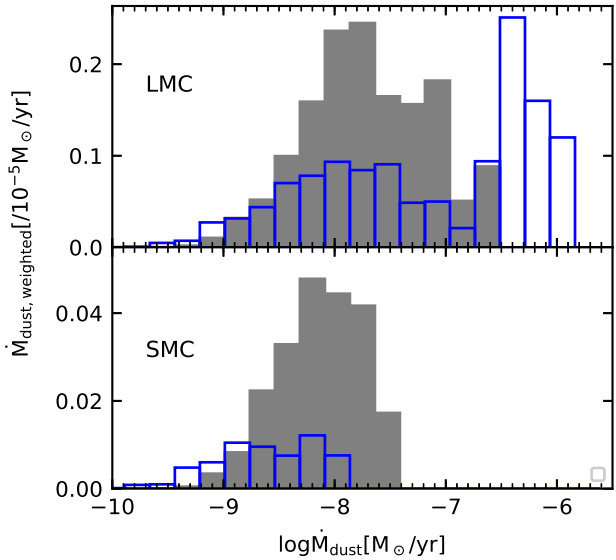}
        \caption{Weighted, normalized distribution functions of the dust production rates computed with the H11 set (grey histograms) and by \citet{Srinivasan16} and \citet{Riebel12} (blue lines) for the carbon stars in the SMC and LMC, respectively.}
        \label{mloss_histo_SR}
        \end{figure}

In Fig.~\ref{mloss_histo_SR} the distributions of the DPRs computed with the H11 set and by \citet{Riebel12}, for the LMC, and by \citet{Srinivasan16}, for the SMC, are compared. The histograms are weighted for the DPRs in each bin. For consistency, we select the sources identified as carbon-rich according to both the GRAMS grids and the photometric classification by \citet{Boyer12}. 
The distributions shown in Fig.~\ref{mloss_histo_SR} are similar for the J1000 case.
For the LMC, the contribution to the total DPR in our analysis comes from stars with a DPR of $-8\lesssim\log{\dot{M}_{\rm dust}}\lesssim-6.5$ with a peak around $\log{\dot{M}_{\rm dust}}\sim-8$, while in \citet{Riebel12} a large fraction of the total DPR is due to stars with $-6.6\lesssim\log{\dot{M}_{\rm dust}}\lesssim-5.7$ that are not found in our analysis. The peak of the distribution derived from \citet{Riebel12} is shifted to values larger than the ones of our study for which $\log{\dot{M}_{\rm dust}}\sim-8$.
For the SMC, the contribution to the total DPR found by \citet{Srinivasan16} is from stars with $-9.4\lesssim\log{\dot{M}_{\rm dust}}\lesssim-8$, while the peak of our distribution is shifted toward larger values of $\log{\dot{M}_{\rm dust}}\sim-8.2$.

The differences found depend on the diverse expansion velocities assumed in \citet{Riebel12}, of 10~km~s$^{-1}$, and \citet{Srinivasan16}, given by equation~\ref{vexp_scaled}, and predicted by our approach (see Fig.~\ref{vel_l_MCs}) and by the different optical constants adopted, as also discussed in Section~\ref{literature}.

In the following we compare our total DPRs with the ones in the literature for the two galaxies.
\subsubsection{SMC}
For the SMC, the DPRs derived for the C- and the X-stars by \citet{Boyer12} and \citet{Srinivasan16} are $\sim3$ times lower than our estimate. This result is also evident from the lower panel of Fig.~\ref{mloss_histo_SR} from which it is clear that our estimate of the DPR is larger than the one in \citet{Srinivasan16} for almost the totality of the stars.

\citet{Matsuura13} based their work on the relation between infrared colours and DPRs based on the SED fitting performed by \citet{Groenewegen07} and on the Surveying the Agents of a Galaxy Evolution (SAGE) data \citep{Meixner06}. In \citet{Groenewegen07} a velocity profile with final wind speed of $10$~km~s$^{-1}$ is adopted. The optical data set for carbon dust is the one by \citet{Rouleau91}. A single value of the grain size of $0.1$~$\mu$m is also assumed, but the optical properties are computed for a continuous distribution of hollow spheres. Our DPR is $\sim 1.6$ times lower than the one derived by \citet{Matsuura13}. This difference is probably related to the different optical properties adopted in \citet{Matsuura13} and in this work. Indeed, in \citet{Nanni18} the DPR estimated by employing the same set of optical constants as in \citet{Matsuura13} provided comparable results.
\subsubsection{LMC}
For the C-stars our DPRs are $\sim 2$ times larger than the one derived by \citet{Riebel12}, and $\sim 4$ times larger than the one by \citet{Srinivasan16}. This discrepancy is probably due to different factors, as discussed in Section~\ref{literature}. For instance, the expansion velocities in \citet{Riebel12} and \citet{Srinivasan16} are lower for a fraction of C-stars (Fig.~\ref{vel_l_MCs}).
Our DPR for X-stars, as well as the total DPR, are in fair good agreement with both \citet{Riebel12} and \citet{Srinivasan16}.
This result can be explained by Fig.~\ref{mloss_histo_SR} from which is clear that the presence of stars with dust production rate around $\log(\dot{M})\approx -6.4$ in \citet{Riebel12} is counterbalanced by the dust production of stars around $\log(\dot{M})\approx -8$ in our analysis.

We find that our estimate is at least $\sim2.4$ times lower than the one provided by \citet{Matsuura09}, who based their analysis on the theoretical estimates of the dust production rates of carbon stars by \citet{Groenewegen07}. The approach is the same as the one adopted for the SMC \citep{Matsuura13}. The difference between our estimate and the lower limit derived by \citet{Matsuura09} can mostly be explained by the diverse optical data sets adopted, as discussed for the SMC. 
The upper limit in the DPR of \citet{Matsuura09} was instead derived from the detection limit in the $[3.6]$ IRAC band, supposing that not all the sources are detected at this wavelength. Thus the upper limit of their DPR might be overestimated in this latter case.

The DPR estimated in this work is also compared with the one by \citet{Dellagli15a}. In this latter work the DPR of the observed stars is estimated according to their colours in some selected colour-magnitude diagrams. Dust growth, coupled with a stationary wind, is described according to \citet{FG06}. This description is similar to the one adopted in our work. 
Despite the similarities of the dust growth prescriptions our DPR is $\sim2.3$ times smaller than their estimate. Such a discrepancy can be due to the diverse set of optical constants and grain sizes obtained and to the different method employed to estimate the DPR.
\begin{table*}
\label{DPR}
\begin{center}
\caption{Total DPRs for the MCs computed for the optical constants in Table~\ref{opt}, and compared with the results found in the literature.}
\begin{tabular}{c l l l l l l}
\hline
       & SMC [$10^{-7} M_{\odot}$~yr$^{-1}$]   &  & &  LMC [$10^{-6} M_{\odot}$~yr$^{-1}$] & & \\
\hline
This work & C-stars  &   X-stars   & Total  &   C-stars  &   X-stars   & Total   \\ 
Number of stars &  1909 &  349       &    2258  & 6907 &  1332       &    8239   \\
J1000&    $3.20\pm0.95$ &   $18.3\pm7.0$ &  $ 21.6\pm8.0$ & $2.90\pm0.85$   &    $13.1\pm3.0$   &    $16.0\pm3.9$ \\
H11&    $4.10\pm1.45$ &   $24.7\pm9.6$ &   $ 28.8\pm11.1$ &  $ 3.38\pm1.04$ &    $16.0\pm4.0$   &    $19.4\pm 5.1$\\
\citet{Boyer12}  &  $\sim 1.2$ &  $\sim6.3$ &  $\sim 7.5$ & - & - & -\\
Number of stars & 1559 & 313  & 1872  & -   &  -    &  -   \\
\citet{Riebel12} & - & -& - & $1.36\pm0.06$  &  $15.7\pm0.6$ &   $17.1\pm0.7$  \\
Number of stars & - & - & - & 6709  &  1340       &   8049    \\
\citet{Srinivasan16}    & $\sim1.2$ &  $\sim6.8$ &  $\sim8.0$ & $\sim0.76$  &  $\sim 12$&   $\sim 12.8$  \\ 
Number of stars & 1652 & 337 & 1989 & 6662  &  1347       &    8009 \\
\citet{Matsuura13}, \citet{Matsuura09}& - & - & $\sim40$ & -  &  - &   $43-100$  \\
\citet{Dellagli15a} &- &- & -&  -   &       -    &   $\sim 40$\\
\hline
\end{tabular}
\end{center}
\end{table*}
\subsection{Carbon stars in the LMC observed with ALMA}\label{subsc:alma}
\begin{figure*}
\includegraphics[trim=0 0 0 0, angle=90, width=0.48\textwidth]{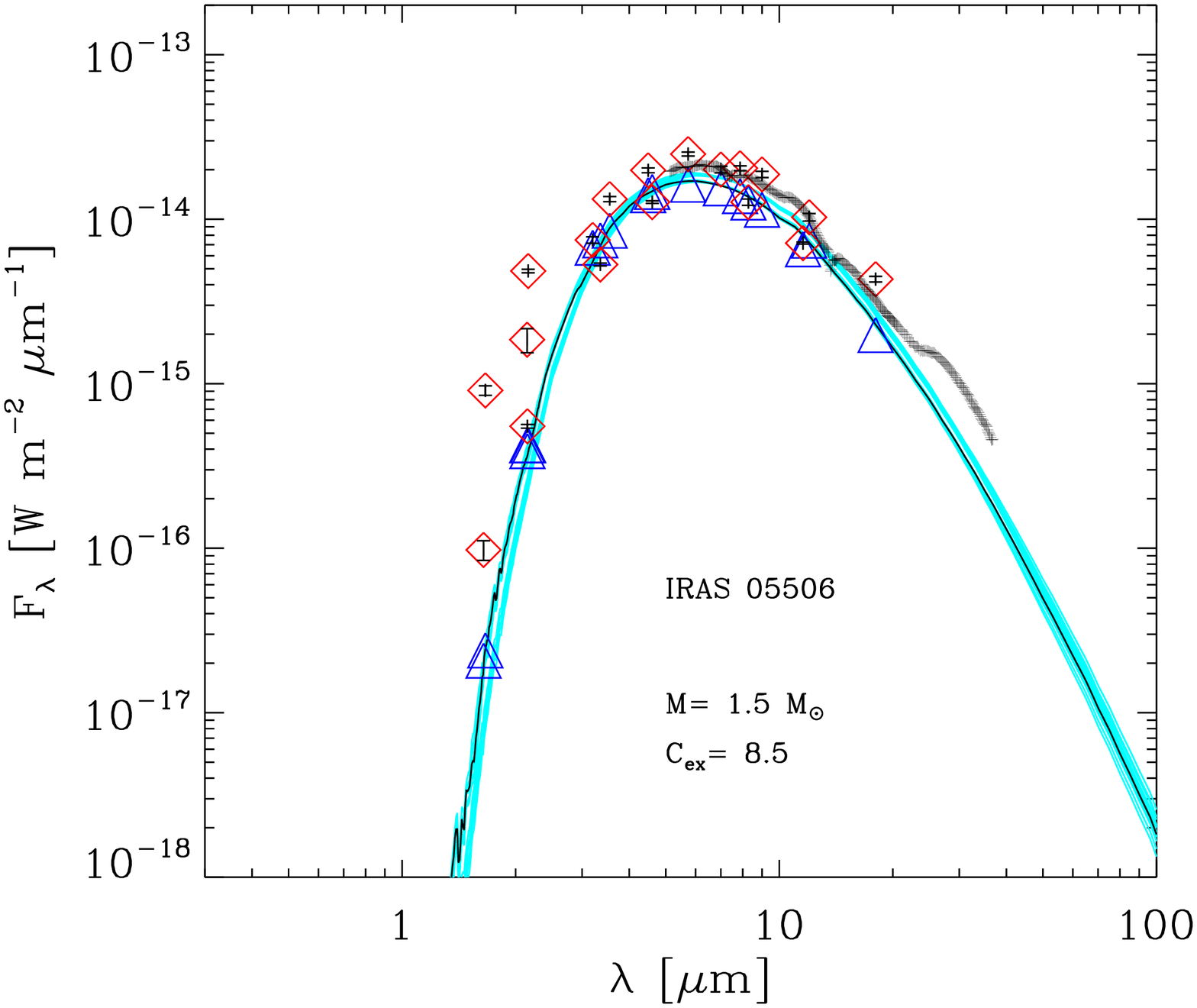}
\includegraphics[trim=0 0 0 0, angle=90, width=0.48\textwidth]{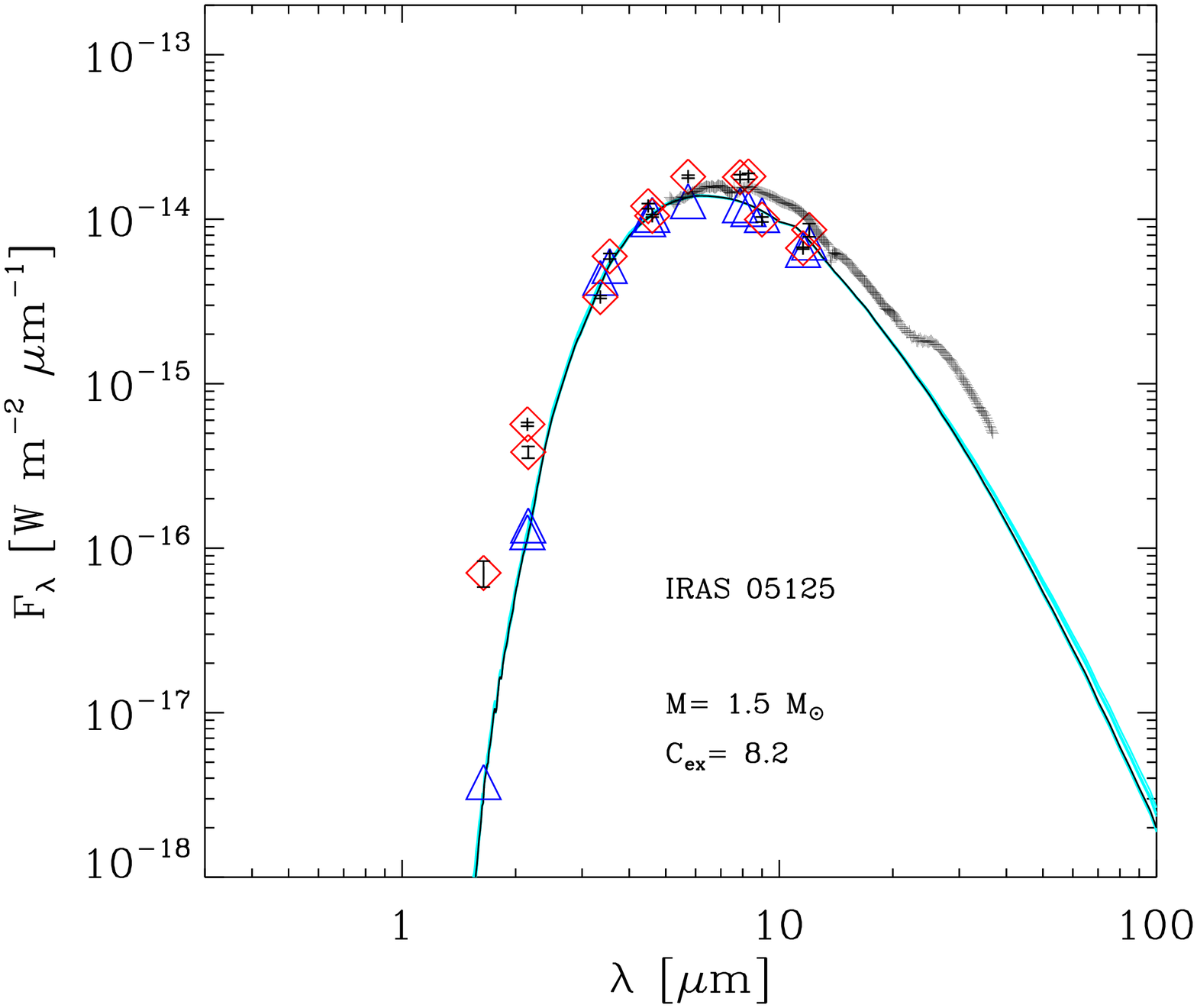}
\includegraphics[trim=0 0 0 0, angle=90, width=0.48\textwidth]{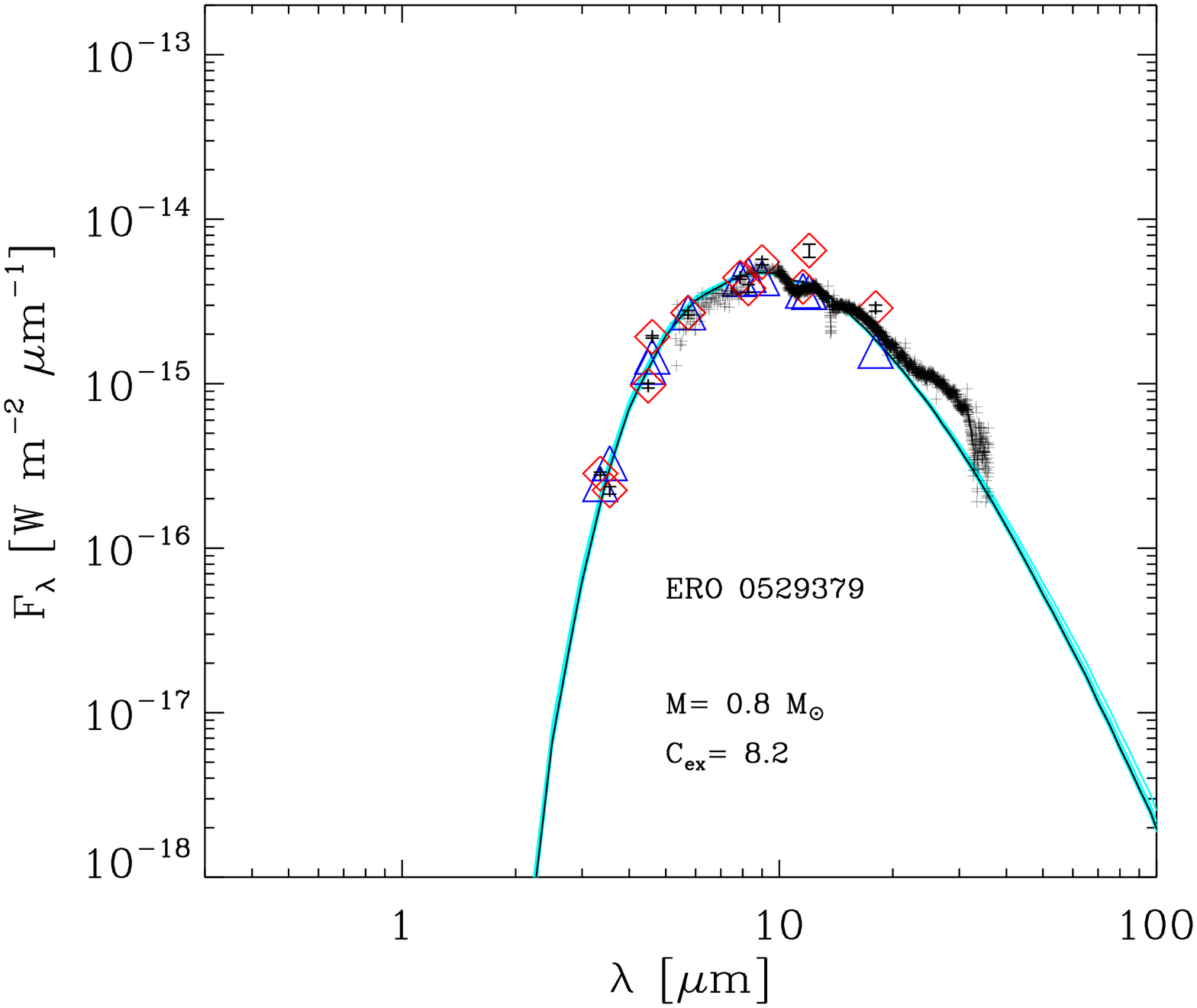}
\includegraphics[trim=0 0 0 0, angle=90, width=0.48\textwidth]{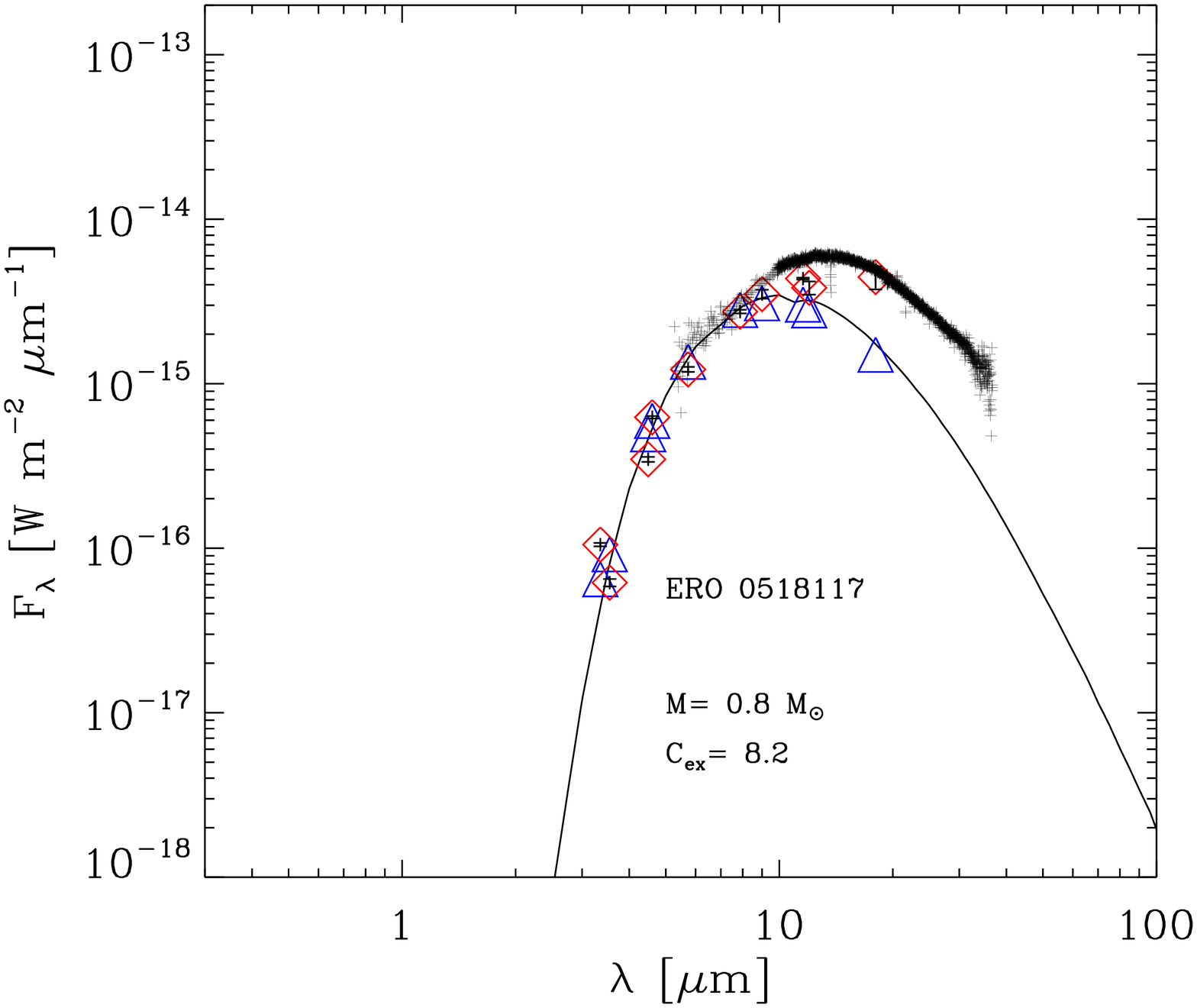}
        \caption{SED fitting for the stars observed with ALMA \citep{Groenewegen16}. The same symbols and line styles as in Fig.~\ref{fit_spectra} are adopted. The value of the current stellar mass and of the carbon-excess are mentioned in each panel.}
        \label{fit_spectra_ALMA}
        \end{figure*}
\begin{figure}
\includegraphics[trim=0 0 0 0, angle=90, width=0.48\textwidth]{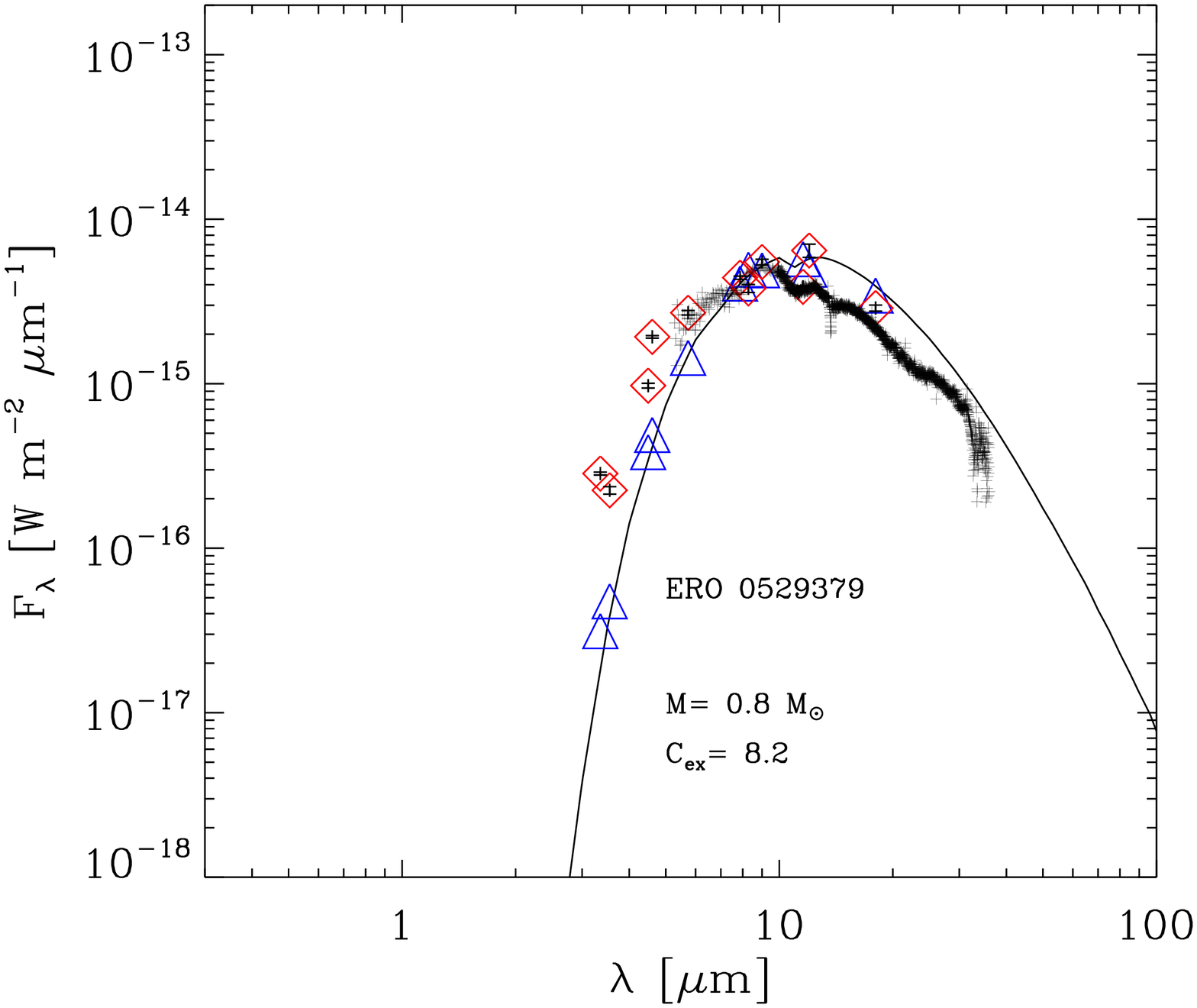}
\includegraphics[trim=0 0 0 0, angle=90, width=0.48\textwidth]{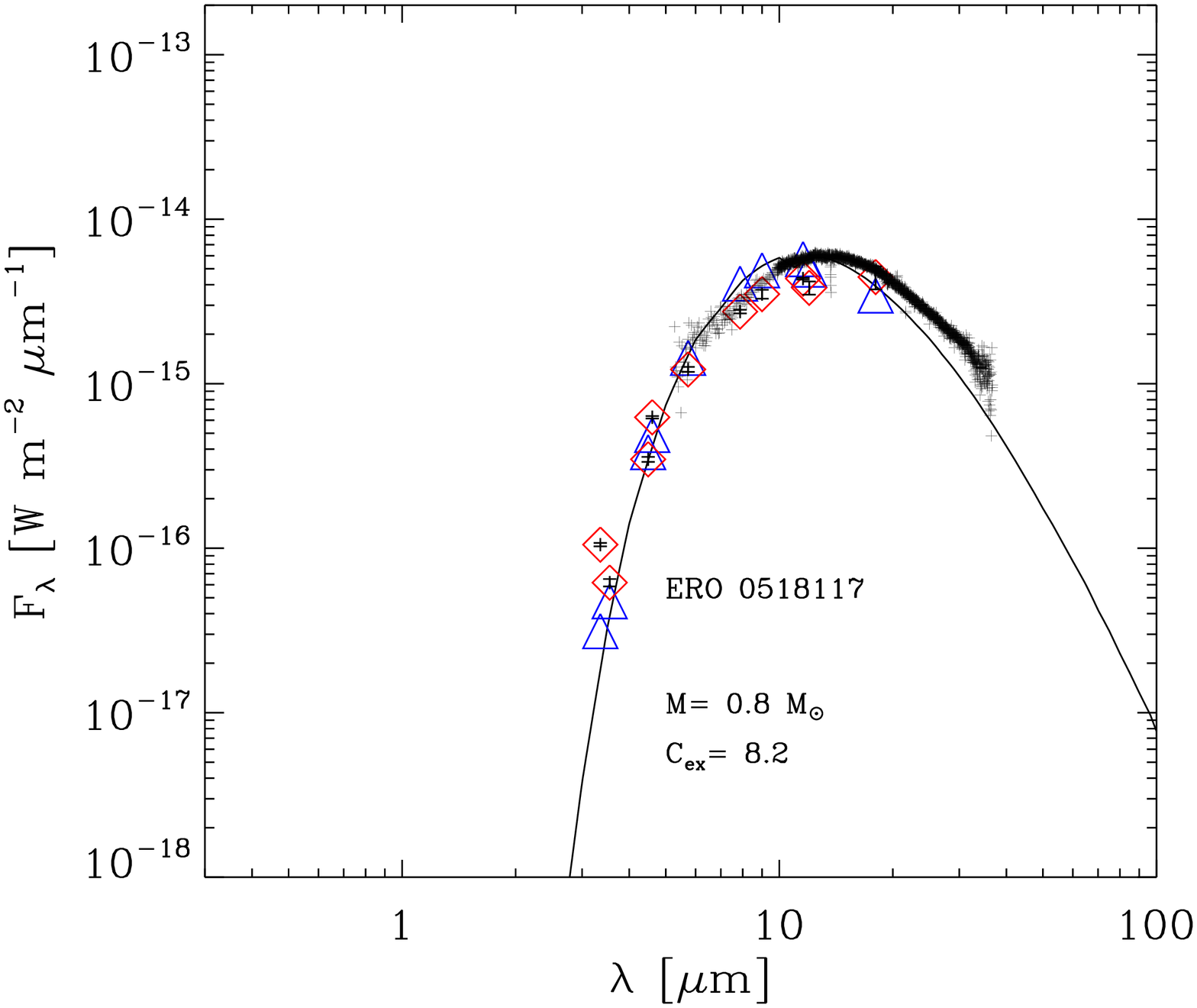}
        \caption{SED fitting for the EROs observed by ALMA \citep{Groenewegen16}. The fit is obtained by employing a large value of the mass-loss rate, $\dot{M}=1\times 10^{-4}$~M$_\odot$~yr$^{-1}$ with $\log(L/L_\odot)=3.9$, $T_{\rm eff}=2500$~K, $M=0.8$~M$_\odot$ and $C_{\rm ex}=8.2$. The same symbols and line styles as in Fig.~\ref{fit_spectra} are adopted.}
        \label{alma_HML}
        \end{figure}
\begin{figure}
\includegraphics[trim=0 0 0 0,  width=0.48\textwidth]{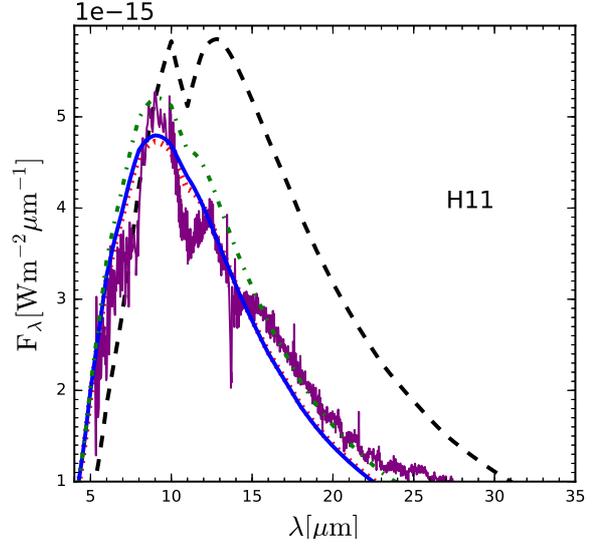}
        \caption{Zoom-in of the ERO 0529379 observed IRS spectrum (purple solid thin line) and of the synthetic spectra obtained from the SED fitting performed with different values of the mass-loss rates and $C_{\rm ex}$ encoded by different line styles. a) Blue solid: $\dot{M}=3.16\times 10^{-5}$~M$_\odot$~yr$^{-1}$, $C_{\rm ex}=8.5$. b) Dotted red: $\dot{M}=4\times 10^{-5}$~M$_\odot$~yr$^{-1}$, $C_{\rm ex}=8.2$. c) Dotted-dashed green: $\dot{M}=4.5\times 10^{-5}$~M$_\odot$~yr$^{-1}$, $C_{\rm ex}=8.2$. d) Dashed black: $\dot{M}=1\times 10^{-4}$~M$_\odot$~yr$^{-1}$, $C_{\rm ex}=8.2$. For further details see Section~\ref{mloss_ERO}.}
        \label{fit_spectra_ERO052}
        \end{figure}
The four carbon stars in the LMC for which the expansion velocities have been derived from CO lines measurements with ALMA by \citet{Groenewegen16} are analyzed in detail. 
For each of the sources the initial and current stellar masses are constrained by employing the TP-AGB tracks from \citet{Marigo13} with initial masses of $1.8$ and $3.0$~M$_\odot$, and metallicity $Z=0.006$.
The same procedure as in \citet{Groenewegen16} is adopted: the current mass of the star is the one of the model in the track with the period and luminosity closer to the observed ones.
The luminosity value adopted is the one derived from this work.
The SED fitting is performed for the different $C_{\rm ex}$ and masses in the grids.
For each star, one example of good SED fitting derived for the H11 set is plotted in Fig.~\ref{fit_spectra_ALMA}. The corresponding values of $C_{\rm ex}$ and current mass from the fitting procedure are mentioned in each panel.
The expansion velocities derived from this analysis are the maximum attained along the CSE, rather than the final ones.
\subsubsection{General properties}
From the SED fitting through the models in our grids, we find that the two optical data sets yield similar values of the outflow expansion velocities and of the other stellar parameters.
A large value of the mass-loss rate of $\sim 3$--$4\times10^{-5}$~M$_{\odot}$~yr$^{-1}$ is derived for all the stars, while the $C_{\rm ex}$ is between $8.2$ and $8.5$. Higher values of the carbon-excess produce too large velocities. A good match with observations is usually obtained for stellar masses in the grids which are close to the ones constrained from the stellar tracks.

As also shown in Fig.~\ref{vel_dmdt_MCs} the carbon stars analysed in this section are among the ones with the largest mass-loss rates in the LMC.
Since these sources are at the extreme end of the mass-loss rate considered in our grids, we decide to test larger values of the mass-loss to be employed in the SED fitting of these sources.
This choice is also justified by the results of \citet{Ventura16}, who studied the $[3.6]$--$[4.5]$ Spitzer colour of the most dust-enshrouded carbon stars in the LMC, and derived a mass-loss rate of $\sim 1.5\times10^{-4}$~M$_\odot$~yr$^{-1}$. In \citet{Ventura16} the stellar evolution is followed by the \textsc{aton} code. Dust growth is based on \citet{FG06} and the radiative transfer calculation is performed with the code \textsc{dusty} \citep{Ivezic97}, similarly to our approach. 
We therefore compute an additional set of spectra for a single value of the effective temperature $T_{\rm eff}=2500$~K and two values of the mass-loss rate, $\dot{M}\sim 1, 1.6 \times 10^{-4}$~M$_\odot$~yr$^{-1}$ and no upper limits for $\tau_1$. The value of $T_{\rm eff}=2500$~K is selected since the most extreme carbon stars are expected to be characterized by low effective temperatures \citep{Marigo08, Ventura16, Marigo17}. Moreover, it is not possible to derive the effective temperature from the SED fitting technique, since the photospheric spectra of these extremely red stars are completely obscured by dust.
The luminosity range is $3.6<\log(L/L_\odot)<4.3$ with the same spacing as in Table~\ref{Table:grid}. The selected luminosities safely include the values estimated for these stars from our standard grids. The spectra are computed for $C_{\rm ex}=8.2, 8.5, 8.7, 9.0$ and for $M=0.8, 1.5, 3.0$~M$_\odot$. The SED fitting is then performed separately for the different combinations of $C_{\rm ex}$ and stellar masses.
With these combinations of parameters, the synthetic spectra are always redder than the observed photometry of the IRAS sources, while acceptable fits are obtained for the Extremely Red objects (EROs) for $\dot{M}\sim 1 \times 10^{-4}$~M$_\odot$~yr$^{-1}$.
This result is not necessarily contradicting the ones by \citet{Ventura16} in which the most extreme stars are expected to be characterized by $3.9<\log(L/L_\odot)<4$ that are closer to the ones estimated for the two EROs.
The fitted SEDs for the two EROs obtained by employing a mass-loss rate of $1\times10^{-4}$~M$_\odot$~yr$^{-1}$ are shown in Fig.~\ref{alma_HML}.
In the following section the main results are discussed.
\subsubsection{The mass-loss rates of the EROs}\label{mloss_ERO}
For the set of models with large $\dot{M}$ the overall SED and wind speed of EROs are reasonably reproduced only for $M=0.8$ M$_{\odot}$, $\dot{M}=1\times10^{-4}$~M$_{\odot}$ and $C_{\rm ex}=8.2$ and $\log(L/L_\odot)=3.9$ for both the optical data sets adopted. The lowest value of $\chi^2_{\rm best}$ is obtained for the H11 set.
The predicted expansion velocity is of $\sim 9.3$~km~s$^{-1}$ for the H11 set and $\sim10$~km~s$^{-1}$ for J1000 which are in reasonably fair agreement with the one observed of $\sim 11$ and $\sim9$~km~s$^{-1}$. For this large value of the input mass-loss rate the SiC feature is in absorption, as also predicted by \citet{Ventura16}. This is very well visible from Fig.~\ref{fit_spectra_ERO052}, where the predicted and observed spectra of ERO 0529379 are shown. From the same figure one can see that the SiC feature is never as deep as in the IRS spectrum when the SED fitting is performed with lower mass-loss rates. The deeper SiC feature is qualitatively in better agreement with the observed IRS spectrum, but the overall SED and spectrum are better fitted by the models with lower mass-loss rates, as can be seen by comparing Fig.~\ref{fit_spectra_ALMA} with Figs.~\ref{alma_HML} and \ref{fit_spectra_ERO052}. In particular, the synthetic photometry obtained with $\dot{M}=1\times10^{-4}$~M$_{\odot}$ is redder than the observed one.
For ERO 0518117 a value of the mass-loss rate as high as $1\times 10^{-4}$~M$_{\odot}$~yr$^{-1}$ improves the fit in the Akari band at $\lambda\sim 18$ $\mu$m with respect to the lower mass-loss case, but the agreement worsens in the IRAC band around $3.6$~$\mu$m, where the predicted absorption is too large. In this case, the SiC feature is predicted to be in absorption in contrast with the observed spectrum.

For these two EROs we refine the fit by computing spectra with mass-loss rates between $4.5$ and $9.0\times 10^{-5}$~M$_{\odot}$~yr$^{-1}$, $M=0.8$ M$_{\odot}$, $T_{\rm eff}=2500$~K, $3.75<\log(L/L_\odot)<3.9$ (with a spacing of $\log(L/L_\odot)=0.05$), $C_{\rm ex}=8.2, 8.5$ and no upper limit for $\tau_1$. 

The combination with $\log(L/L_\odot)=3.75$, $\dot{M}\sim4.5\times10^{-5}$~M$_{\odot}$~yr$^{-1}$ and $C_{\rm ex}=8.2$ satisfactorily reproduces the overall SED as well as the expansion velocity of ERO 0529379, as also shown in Fig.~\ref{fit_spectra_ERO052}.
For the H11 set, the quality of the fit is comparable with the one obtained with the lower mass-loss rates, while for the J1000 the fit is worse than in the low mass-loss rate case. 
Nevertheless, also for this combination of parameters the SiC feature does not appear as deep as in the observed spectrum (see Fig.~\ref{fit_spectra_ERO052}). \citet{Sloan16} suggested that these sources might have non-spherical geometry that can affect the appearance of the spectrum.
For ERO 0518117 the fit is improved with respect to the standard grids by adopting the following combination of parameters: $\log(L/L_\odot)=3.75$, $\dot{M}\sim6.3\times10^{-5}$~M$_{\odot}$~yr$^{-1}$ and $C_{\rm ex}=8.2$.
The expansion velocity of $\sim 8.87$~km~s$^{-1}$ is also reasonably reproduced, with a value of $\sim 9.1$ and $\sim 8.3$~km~s$^{-1}$ for the J1000 and H11 data set, respectively. The best fit is obtained with the H11 data set.

From the above discussion it is possible to conclude that a mass-loss rate of $\sim 1\times 10^{-4}$~M$_\odot$~yr$^{-1}$ can only be considered as a generous upper limit of the mass-loss rate of the EROs analysed here. Indeed, these sources are usually better fitted by mass-loss rates $\sim3$ times lower than the ones predicted by \citet{Ventura16}.
How to obtain such large values of the mass-loss rates and similar expansion velocities for this combination of parameters represents an open issue \citep{Mattsson10, Eriksson14} that is discussed in Section~\ref{cfr_model}.

\subsubsection{Comparison with hydrodynamic calculations and stellar evolution models}\label{cfr_model}
A summary of our best-fitting parameters derived for each of the ALMA sources is provided in Table~\ref{alma} together with the model selected along the tracks to match the observed period and luminosity.

The values of the carbon-excess required to reproduce the expansion velocities observed ($C_{\rm ex}=8.2, 8.5$) are partly in agreement with the results of the hydrodynamic calculations \citep{Mattsson10,Nowotny13,Eriksson14}. In particular, Fig.~4 of \citet{Eriksson14} shows that expansion velocities between $10$ and $20$~km~s$^{-1}$ are achieved for $C_{\rm ex}=8.5$. For $C_{\rm ex}=8.2$, the typical expansion velocity obtained by \citet{Eriksson14} is $\lesssim4$~km~s$^{-1}$. This value is lower than the ones derived by our analysis and listed in Table~\ref{alma}. The discrepancy in the predicted expansion velocities can be ascribed to the chosen optical data sets adopted in \citet{Mattsson10} and \citet{Eriksson14} and to the different grain sizes predicted. Indeed, in \citet{Mattsson10} and \citet{Eriksson14} the optical constants by \citet{Rouleau91} are employed. This choice of the optical data set produces lower expansion velocity for the outflow with respect to the H11 and J1000 sets \citep{Nanni18}. A dependence on the final grain size for the same set of optical constants is also shown in \citet{Nanni18}.
The large values of mass-loss rates of $3-6\times10^{-5}$~M$_\odot$~yr$^{-1}$ derived from our SED fitting procedure are never achieved in the hydrodynamic simulations in which the mass-loss rate is usually below $\sim10^{-5}$~M$_\odot$~yr$^{-1}$ . 

The models selected along the stellar tracks for each star have mass-loss rates systematically lower, by a factor between $1.5$ and $5$, than the ones predicted by our SED fitting. The value of the carbon-excess is instead usually larger than the one we derive for a current mass usually compatible with the one of the tracks.
The only exception is represented by IRAS 05506 -- 7053 (IRAS 05506) for which the carbon-excess in the track is similar to the one obtained from the SED fit.
For the EROs the mass-loss rates of the models selected along the tracks are $\sim 8$--$8.6\times10^{-6}$~M$_\odot$~yr$^{-1}$. 
According to the analysis presented here, such values of the mass-loss rate are not able to reproduce the very extreme colours of the EROs.
As shown in Fig.~10 of \citet{Eriksson14} and in Fig.~7 of \citet{Nowotny13} this result is confirmed by the hydrodynamic calculations, where the value of the $\jks$ colour attained for $\dot{M}\sim10^{-5}$~M$_\odot$~yr$^{-1}$ is $\sim 5$~mag. 

Alternative ways to obtain redder spectra for a given mass-loss rate would require an increase in the carbon-excess and/or of the dust condensation efficiency in dust growth models. This would allow to condense the same amount of dust with a lower mass-loss rate. 
Values of the carbon excess up to $C_{\rm ex}=9$ are tested for the stars discussed here. Such large values of the $C_{\rm ex}$ always produce too high velocities for a given stellar mass, while, at the same time, the value of the mass-loss rate derived from the fit is never as low as $\sim (8-8.6)\times10^{-6}$~M$_\odot$~yr$^{-1}$. 

It is also problematic to change the underlying dust growth prescriptions in order to increase the amount of amC dust condensed.
A larger amount of amC dust might be produced by more efficiently accreting the available C$_2$H$_2$ molecules onto amC dust grains. 
However, both in hydrodynamic simulations and in the calculations presented here the sticking coefficient of amC dust, $\alpha_{\rm amC}$, is already set to its maximum value ($\alpha_{\rm amC}=1$ in equation~\ref{Jgr}). \citet{Mattsson10} tested a lower value of the sticking coefficient, $\alpha_{\rm amC}=0.5$, in addition to the standard case with $\alpha_{\rm amC}=1$ and found that similar results are obtained in the two cases.
The same values of the sticking coefficient have been employed in \citet{Nanni13} who found that the total amC dust yields from the TP-AGB phase is only $\sim5$ per cent different in the two cases. 
The effect of condensing amC dust at gas temperatures higher ($1300$~K) than the standard assumption of $1100$~K based on \citet{Cherchneff92}, has also been tested in \citet{Nanni13}. The increase of the condensation temperature slightly reduces the amount of amC dust formed, thus not solving the issue. 

On the basis of the tests performed in \citet{Mattsson10} and \citet{Nanni13} it is possible to conclude that large variation of the amC dust mass is difficult to be achieved by only changing the input parameters in the dust condensation prescriptions, and that a large mass-loss rate is probably needed in order to reproduce the photometry of the most obscured stars, as long as symmetric mass-loss is assumed.

\subsubsection{Comparison with Groenewegen et al. 2016}
The gas-to-dust ratios and mass-loss of the selected stars are compared with the ones derived by \citet{Groenewegen16}, provided in Table~\ref{alma}.
Our estimates of the mass-loss rates are in fair agreement with the ones by \citet{Groenewegen16} for IRAS 05125 and ERO 0529379 (see Section~\ref{mloss_ERO}). For IRAS 05506 our mass-loss rate is $\sim2$ times larger than the one by \citet{Groenewegen16}. 
For ERO 0518117 our mass-loss rate is $\sim1.6$--$1.8$ times larger than the one by \citet{Groenewegen16}.

Our estimate of the gas-to-dust ratio is always lower than the one derived by \citet{Groenewegen16} in which this quantity is scaled with the velocity obtained from the wind dynamics in the code \textsc{dusty}. 
The discrepancy found is possibly due to the treatment of wind dynamics in the \textsc{dusty} code. 
Indeed, the outflow is forced to accelerate at the dust condensation zone, determined by the dust temperature. This latter quantity is however an input parameter of \textsc{dusty} that is therefore not derived by consistently computing grain growth. The wind velocity from \textsc{dusty} will be in general different from the one derived by our method, and this will affect the estimate of the gas-to-dust ratio (and mass-loss rate). 
Another source of difference between this analysis and in the one by \citet{Groenewegen16} is related to the choices of the optical data sets and of the grain size and dust-density profile. In our treatment we assume the CSE to be spherically symmetric, however deviations from the spherical symmetry for the reddest source have been suggested \citep{Sloan16}. 
\begin{table*}
\begin{center}
\caption{Observed and predicted properties for the ALMA sample of carbon stars \citep{Groenewegen16}. For each star, the observed quantities are listed in the first line, together with the mass-loss rate estimated by \citet{Groenewegen16} through the SED fitting technique. In the second line the period, luminosity, initial and current mass and the carbon-excess from the TP-AGB tracks are provided \citep{Marigo13}. In the last two lines the quantities derived from the SED fitting for the J1000 set (third line) and the H11 one (fourth line) are given.}
\label{alma}
\begin{tabular}{c c c c c c c c c c}
\hline
 IRAS name & Identifier & $v_{\rm exp}$ [km~s$^{-1}$] & P [days] & $\log(L/L_\odot) $ &$M_{\rm i}$ [$M_{\odot}$] & $M$ [$M_{\odot}$]  & $C_{\rm ex}$ &   $\dot{M}/10^{-5}$ [$M_{\odot}$~yr$^{-1}$] & $\Psi$  \\
\hline
05506 -- 7053 & IRAS 05506 & $23.63\pm 0.42$ & $1026$ &  $4.25$    & - & - & - & $\sim 1.6$ & $133$ \\ 
Track     &      & -           &  $1026$   & $4.17$ & $3$ & $1.85$ & $\sim 8.50$ & $\sim2$    & -  \\
J1000     &  & $21.46$--$24.05$ & -   & $4.12$ &  - & $1.5$--$3.0$ & $8.5$    & $\sim2.97$--$3.32$ & $416$--$422$   \\
H11    &  & $22.37$--$25.32$ & - & $4.12$--$4.13$ & -  & $1.5$--$3.0$ &  $8.5$   & $\sim2.92$--$3.10$  & $359$--$360$ \\
05125 -- 7035 & IRAS 05125 & $11.77\pm 0.15$ & $1115$ & $4.19$   & -  &     -  & - &$\sim 4$ & $541$ \\
Track      &     &  -           & $1116$ & $4.03$ & $3$ & $1.30$ & $\sim 8.56$ & $\sim 1.47$ & -\\   
J1000    &  & $11.39$--$14.63$ & -   & $4.05$--$4.07$ & -  & $1.5$--$3.0$   & $8.2$  &   $\sim 3.39$--$3.98$ & $705$--$722$  \\
H11      &  & $11.04$--$14.26$ & -   & $4.05$ &  - & $1.5$--$3.0$   & $8.2$ &    $\sim 3.16$--$3.98$ & $662$--$665$ \\
05305 -- 7251 & ERO 0529379 & $11.04\pm 0.41$ & $1076$ &  $3.73$    &    -     &    - & -    &  $\sim 4.5$ & $504$\\
Track       &    &  -           &     $1067$   & $3.78$     &   $1.8$ &  $0.74$ &  $\sim8.64$ &$\sim 0.82$ & - \\ 
J1000  &  & $9.63$--$15.27$ & - & $3.70$ & -  &  $0.8$  &  $8.2$--$8.5$ &     $\sim 2.81$--$3.69$ & $359$--$667$ \\
H11 &  & $8.88$--$13.91$ & - & $3.70$--$3.75$ & -  & $0.8$  & $8.2$--$8.5$ &    $\sim 3.16$--$4.5$ & $353$--$619$ \\
 05187 -- 7033 & ERO 0518117       &  $8.87\pm 0.52$ &  $1107$ & $3.97$   &     -   & -  &  - & $\sim 3.6$ & $142$ \\
Track          &   &    -             &  $1107$     & $3.80$       & $1.8$ & $0.83$ & $\sim 8.64$ &  $0.86$ & -\\   
J1000 &  & $9.1$  & -  & $3.75$  &    -    &  $0.8$ & $8.2$    & $\sim 5.6$ & $659$             \\
H11  &  & $8.3$& -    & $3.75$  &    -    &  $0.8$ & $8.2$     &  $\sim 6.3$ & $557$             \\
\hline
\end{tabular}
\end{center}
\end{table*}
\section{Caveats and uncertainties}\label{sec:caveats}
We here discuss the main sources of uncertainty and how our assumptions in the dust growth calculations affect the results.
\begin{itemize}
\item \textit{Stationary wind}. We adopt a description of stationary outflow assuming an input mass-loss rate.
This approach is computationally light and thus suitable to produce number of spectra much larger than the one computed by means of hydrodynamic codes \citep{Mattsson10, Eriksson14,Bladh19}. 
In our approach a time-dependent description of the wind that includes pulsations and the density enhancement due to shocks at the dust condensation region is missing. To check the possible differences we compare our gas-to-dust ratio with the ones computed with hydrodynamic simulations in a similar range of carbon-excess and mass-loss rates \citep{Mattsson10, Eriksson14}. Interestingly, we find that the values we derived are comparable with the ones of more detailed models.

Drift velocity of the dust grains with respect to the gas is not taken into account in our calculations. When drift velocity is considered, the dust-to-gas ratio increases of $30$ per cent for a given mass-loss rate \citep{Krueger97}. The final wind speed of the gas is instead $\sim30$ per cent lower for the low mass-loss rates and almost unchanged for the large mass-loss rates.
In order to obtain the same $\tau_\lambda$ as the one obtained in our framework (equation~\ref{tau_lambda}) a lower value of the mass-loss rate is required. This implies that our assumptions are likely to overestimate the mass-loss rate. On the other hand, we expect that the DPR will be larger only for the lower mass-loss rates, for which the gas expansion velocities are lower. Thus we do not expect dramatic variations in the total DPR.
\item\textit{Sticking coefficient of amC dust and $v_{\rm i}$}. 
The sticking coefficient for amC dust determines the condensation efficiency of this dust species. In \citet{Nanni13} two different values of the sticking coefficient for amC dust ($\alpha_{\rm amC}=0.5, 1$) have been shown to well reproduce the observed trend between the wind speed and mass-loss rate in our Galaxy.
On the other hand, the initial expansion velocity is selected to be $v_{\rm i}=4$~km~s$^{-1}$ in the present calculations. This model parameter determines the initial density of the outflow (see equation~\ref{dens}). Variations of $\alpha_{\rm amC}$ and $v_{\rm i}$ are also expected to change the optical depth for a given set of input stellar parameters. 
In order to check the sensitivity of the results on the input parameters, we vary $\alpha_{\rm amC}$ and $v_{\rm i}$ separately, by computing two grids of models with $\alpha_{\rm amC}=0.5$ and $v_{\rm i}=1$~km~s$^{-1}$. We consider the H11 data set in the following range of stellar parameters: $-5\le\log\dot{M}\le-4.5$,  $3.6\le\log{L}\le4.1$, $2500\le T_{\rm eff}\le 2800$~K. The same stellar parameters as in Table~\ref{Table:grid} has been adopted for the other quantities. 
We select $20$ stars in the LMC lying in the aforementioned range of parameters when fitted with the standard grids of spectra ($\alpha_{\rm amC}=1$, $v_{\rm i}=4$~km~s$^{-1}$) and we perform the SED fitting with the new grids.
The $\chi^2_{\rm best}$ is similar to the standard case for $\alpha_{\rm amC}=0.5$, while it is generally worse (but still acceptable) for $v_{\rm i}=1$~km~s$^{-1}$.

The variation of the amC dust sticking coefficient produce a total decrease of this species of about $\sim 10$ per cent, and a corresponding increasing of the SiC of the $\sim7$ per cent in mass. These variations are within the uncertainties found from the SED fitting procedure.
The expansion velocity and DPR of individual sources is typically between $\sim20 - 40$ per cent and $\sim10 - 30$ per cent lower, respectively for $\alpha_{\rm amC}=0.5$, while the mass-loss rate derived is usually comparable or larger.

The total DPR obtained by assuming $v_{\rm i}=1$~km~s$^{-1}$ is $\sim 30$ per cent lower than the standard case. Such a variation is close to the uncertainty of $\sim 25$ per cent given in Table \ref{DPR}.
The final wind speed is $\sim 20$ per cent smaller than in the standard case, as well as the DPR and mass-loss rates of the individual stars (by a factor $\sim15-45$ and $\sim10$ per cent, respectively).

We finally emphasise that we expect only minor variation in the total dust budget due to changing in the value of $\alpha_{\rm SiC}$ that mostly affects the abundance of SiC dust and the $11.3$~$\mu$m feature.
\item\textit{Deviation from homogeneity in the CSEs}. Some of the observed stars are likely to be characterised by a clumpy structure \citep{Sloan16}. \citet{VandeSande18} showed that the optical depth is lower for clumpy mediums than in the homogeneous case for the same value of mass-loss rate. Thus, in case the medium is non-homogeneous, the DPR is probably underestimated under our assumptions.
\item\textit{Stellar variability}. An additional source of uncertainties is related to the stellar variability of individual objects. 
In \citet{Riebel12} and \citet{Srinivasan16} the initial photometric error is increased in order to account for the variability from the U to the $K_{\rm s}$ band. The quantity added to the photometric error is estimated for the V band, and is assumed to be the same for all the bands from U to $K_{\rm s}$.
Since larger amplitude variations are found for the shortest wavelengths \citep{Nowotny11}, this procedure underestimate the error at shorter wavelengths and overestimate it at the longer wavelengths. We simulate the effect of different variability amplitudes at different wavelengths in the following way.
For $4932$ carbon stars in the LMC for which the variability amplitude is available, we follow the same procedure described in Section $2.2$ of \citet{Riebel12}, but we add to the photometric error a term proportional to the variability in the different bands as in Fig. 9 of \citet{Nowotny11}. The value for the U band is extrapolated.
The ability of fitting the sources (with the H11 set) is generally not altered, with the exception of few cases.
The total DPR derived is different from the standard case only by few per cents. The DPR of the individual stars follows approximately the same distributions as in the standard case, while the peaks of the gas-to-dust ratios and of mass-loss rates correspond to half of the previously derived value. These two quantities have a typical uncertainty of $\sim 40$ and $\sim 25$ per cent, respectively. Most of these stars are characterised by low values of the mass-loss rates ($\sim 1-2 \times 10^{-7}$ M$_\odot$~yr$^{-1}$), that also explains the reason for the modest variation in the total dust budget. The peak of the wind speed distribution is shifted towards larger values ($\Delta v_{\rm max}\sim12$~km~s$^{-1}$), while the typical uncertainty is between $\sim 5-9$~km~s$^{-1}$.
\end{itemize}
\section{Summary and conclusions}
In this work, we performed the SED fitting technique over the spectra computed for two grids of stellar parameters for all the carbon stars in the MCs. Each of the grids of spectra is computed for a combination of optical constants and grain sizes of amC dust that is able to reproduce different colour--colour diagrams in the infrared as well as specific observations in the optical bands performed with \textit{Gaia} DR2 \citep{Nanni16, Nanni19}. Dust growth is coupled with a stationary and spherically symmetric wind \citep{FG06, Nanni13}. The spectra reprocessed by dust are computed with the code \textsc{more of dusty} \citep{Groenewegen09, Ivezic97} that takes as input some of the output quantities obtained from the dust growth code. This approach allows us to predict the gas-to-dust ratios, dust chemistry and outflow expansion velocity as a function of the stellar parameters. 
The most relevant findings of this investigation are summarised in the following.
\begin{itemize}
\item \textit{Mass-loss rates}. 
The mass-loss rates derived for the sample in the SMC are lower than the ones in the LMC. 
Such a difference might be due to the different luminosity function of the SMC sources with respect to the LMC ones.  
\item \textit{Gas-to-dust ratios}. 
The gas-to-dust ratio derived covers a large range of values as a function of the mass-loss rate. A larger condensation efficiency is found for larger mass-loss rates. The typical gas-to-dust ratio of the X-stars is $\approx 700$ for both the MCs. 
A fraction of X-stars in the LMC shows a dust production efficiency larger than the one in the SMC. 
The minimum values achieved in the two galaxies are $\sim 100$, for the LMC, and $\sim160$--$200$, for the SMC. 
\item \textit{Outflow expansion velocity}. A fraction of C-stars surrounded by a low amount of dust exhibit expansion velocities $<10$~km~s$^{-1}$, in agreement with the scaling relation assumed in different works \citep{vanLoon00, Boyer12, Srinivasan16}. However, for a large fraction of stars the predicted expansion velocities are much larger ($\sim 30$~km~s$^{-1}$).
The maximum wind speeds attained (between $\sim 50$ and $\sim 60$~km~s$^{-1}$) are not observed in our Galaxy, and should thus be confirmed by future direct observations. In case the wind speed predicted in our approach is overestimated, the DPR would be also inflated.
\item \textit{Total dust production rate.} 
Our total DPRs are $\sim 1.77\pm 0.45\times10^{-5}$~M$_\odot$~yr$^{-1}$, for the LMC, and $\sim 2.52\pm 0.96 \times 10^{-6}$~M$_\odot$~yr$^{-1}$, for the SMC.
For the LMC, our DPRs are compatible with the ones provided by \citet{Riebel12} and \citet{Srinivasan16}, even though we find larger values for the C-stars.
Our estimate is instead more than $\sim2$ times smaller than the ones by \citet{Matsuura09} and \citet{Dellagli15a}. 
For the SMC, our DPRs are $\sim3$ times larger than the ones by \citet{Boyer12, Srinivasan16} and $\sim 1.8$ times smaller than the one by \citet{Matsuura13}. This latter result is probably opacity-dependent, since in \citet{Nanni18} a good agreement was found for similar optical data sets.
\item \textit{ALMA carbon stars}. The stars observed by ALMA by \citet{Groenewegen16}, are analyzed in detail. 
For the IRAS sources the estimated mass-loss rate is $\sim (3$--$4)\times10^{-5}$~M$_\odot$~yr$^{-1}$, while higher values, $\sim (4.5-6.3)\times 10^{-5}$~M$_\odot$~yr$^{-1}$ are derived for the EROs. The IRS spectrum of  ERO 0529379 shows a deep SiC absorption feature which shows up in our models only for large values of the mass-loss rate, $\dot{M}=1\times10^{-4}$~M$_\odot$~yr$^{-1}$, that is however not reproducing the overall SED and spectrum. Non-spherical geometry might play a role.

Our estimated mass-loss rates are in fair agreement with the ones by \citet{Groenewegen16}, however our values of the gas-to-dust ratio are always larger. 
This discrepancy might also depend on the assumptions related to the wind dynamics in the code \textsc{dusty} that always forces the outflow acceleration under the assumption that mass-loss is occurring. 
\item \textit{Caveats and uncertainties.} We test the sensitivity of our results due to different assumptions in our calculations (i.e. sticking coefficient of amC dust and initial wind speed). Reasonable variations in the input parameters seem not significantly affecting our estimate of the total DPR.
\end{itemize}
Our grids of models and spectra are publicly available together with the fitted sources at \url{https://ambrananni085.wixsite.com/ambrananni/online-data-1}, as well as in the SED-fitting python package for fitting evolved stars \url{https://github.com/s-goldman/Dusty-Evolved-Star-Kit}.
\subsection*{Acknowledgements}
This work was supported by the ERC Consolidator Grant funding scheme
({\em project STARKEY}, G.A. n.~615604) and by the Centre National d'{\'E}tudes Spatiales (CNES).
AN and JvL acknowledge an award from the Sir John Mason Academic Trust in support of theoretical investigations of dusty winds and evolved
star populations.
We thank Professor Paola Marigo for the useful discussion and for kindly providing the stellar tracks used in Section~\ref{subsc:alma}. We thank the referee for his/her precious comments that significantly improved the manuscript.
This publication makes use of data products from the Two Micron All Sky Survey, which is a joint project of the University of Massachusetts and the Infrared Processing and Analysis Center/California Institute of Technology, funded
by the National Aeronautics and Space Administration (NASA) and
the National Science Foundation. 
This work is based in part on observations made with the Spitzer Space Telescope, which is operated by the Jet Propulsion Laboratory, California Institute of Technology under contract with NASA. This publication makes use of data
products from the Wide-field Infrared Survey Explorer, which
is a joint project of the University of California, Los Angeles, and
the Jet Propulsion Laboratory/California Institute of Technology, funded by NASA. This research makes use of observations
with AKARI, a Japan Aerospace Exploration Agency project
with the participation of European Space Agency.
This research made use of
Astropy, a community-developed core Python package for
Astronomy \citep{Astropy13,Astropy18} and
matplotlib, a Python library for publication quality graphics \citep{Hunter07}.
\bibliographystyle{mnras}
\bibliography{nanni}
\bsp
\label{lastpage}
\end{document}